\documentclass[twocolumn,prl,amsmath,amssymb,floatfix,reprint,square,comma,numbers]{revtex4-1}
\usepackage{graphicx}
\usepackage{ulem}
\usepackage{amstext}
\usepackage{makeidx}
\usepackage[colorlinks=true,linkcolor=blue]{hyperref}
\usepackage{color,soul}
\usepackage{amsmath}
\usepackage{braket}

\begin{document}

\title{10-qubit entanglement and parallel logic operations
with a superconducting circuit}

\author{Chao Song$^{1,2}$}
\thanks{C. S. and K. X. contributed equally to this work.}
\author{Kai Xu$^{1,2}$}
\thanks{C. S. and K. X. contributed equally to this work.}
\author{Wuxin Liu$^1$, Chuiping Yang$^3$}
\author{Shi-Biao Zheng$^4$}
\email{t96034@fzu.edu.cn}
\author{Hui Deng$^5$, \mbox{Qiwei Xie$^6$}, Keqiang Huang$^{5,8}$, Qiujiang Guo$^1$, Libo Zhang$^1$, Pengfei Zhang$^1$, Da Xu$^1$, \mbox{Dongning Zheng$^{5,8}$}}
\author{Xiaobo Zhu$^{2,9}$}
\email{xbzhu16@ustc.edu.cn}
\author{H. Wang$^{1,2}$}
\email{hhwang@zju.edu.cn}
\author{Y.-A. Chen$^{2,9}$, C.-Y. Lu$^{2,9}$, Siyuan Han$^7$}
\author{Jian-Wei Pan$^{2,9}$}
\affiliation{$^1$\mbox{Department of Physics, Zhejiang University, Hangzhou, Zhejiang 310027, China},
$^2$CAS Center for Excellence and Synergetic Innovation Centre in Quantum Information and Quantum Physics, University of Science and Technology of China, Hefei, Anhui 230026, China,
$^3$\mbox{Department of Physics, Hangzhou Normal University,~Hangzhou,~Zhejiang 310036, China},
$^4$Fujian Key Laboratory of Quantum Information and Quantum Optics, \mbox{College of Physics and Information Engineering, Fuzhou University, Fuzhou, Fujian 350116, China},
$^5$\mbox{Institute of Physics, Chinese Academy of Sciences, Beijing 100190, China},
$^6$\mbox{Institute of Automation, Chinese Academy of Sciences, Beijing 100190, China},
$^7$\mbox{Department of Physics and Astronomy, University of Kansas, Lawrence, Kansas 66045, USA},
$^8$\mbox{School of Physical Sciences, University of Chinese Academy of Sciences, Beijing 100049, China},
$^9$ Shanghai Branch, National Laboratory for Physical Sciences at Microscale and Department of Modern Physics,
University of Science and Technology of China, Shanghai 201315, China}

\begin{abstract}
Here we report on the production and tomography of genuinely entangled Greenberger-Horne-Zeilinger states 
with up to 10 qubits connecting to a bus resonator in a superconducting circuit, where 
the resonator-mediated qubit-qubit interactions are used to controllably 
entangle multiple qubits and to operate on different pairs of qubits in parallel. 
The resulting 10-qubit density matrix is probed by quantum state tomography, with a fidelity of $0.668\pm 0.025$.
Our results demonstrate the largest entanglement created 
so far in solid-state architectures, and pave the way to large-scale quantum computation.
\end{abstract}

\maketitle

Entanglement is one of the most counter-intuitive features of quantum mechanics. 
The creation of increasingly large number of maximally entangled quantum bits (qubits) is central
for measurement-based quantum computation~\cite{Raussendorf2001},
quantum error correction~\cite{Calderbank1996,Knill2005}, quantum
simulation~\cite{Lloyd1996}, and foundational studies of nonlocality~\cite{Greenberger1990,Ansmann2009}
and quantum-to-classical transition~\cite{Leggett2008}. 
A significant experimental challenge for engineering multiqubit
entanglement~\cite{Wang2016,Monz2011,Barends2014} has been noise control~\cite{Wei2006,Matsuo2006}.
With solid-state platforms, the largest
number of entangled qubits reported so far is five~\cite{Barends2014},
and further scaling up would be difficult as constrained
by the qubit coherence and the employed sequential-gate method.

Superconducting circuits are a promising solid-state platform for quantum
state manipulation and quantum computing owing to the microfabrication
technology scalability, individual qubit addressability, and ever-increasing
qubit coherence time~\cite{You2011}. The past decade has witnessed significant progresses
in quantum information processing and entanglement engineering with
superconducting qubits: preparation of three- and four-qubit entangled states~\cite{DiCarlo2010,Neeley2010,Zhong2016,Paik2016},
demonstration of elementary quantum algorithms~\cite{DiCarlo2009,Mariantoni2011}, realization of
three-qubit Toffoli gates and quantum error correction~\cite{Fedorov2012,Reed2012,Riste2013,Riste2015,Takita2016}. In
particular, a recent experiment has achieved a two-qubit controlled-phase
gate with a fidelity above 99 percent with a superconducting quantum
processor~\cite{Barends2014}, where five transmon qubits with nearest-neighbor coupling
are arranged in a linear array. Based on this gate, a 5-qubit Greenberger-Horne-Zeilinger (GHZ) state
was produced step by step; the number of entangled qubits is increased by one at a time.
With a similar architecture consisting of 9 qubits, digitized Trotter steps were
used to emulate the adiabatic change 
of the system Hamiltonian 
that encodes a computational problem~\cite{Barends2016},
where the digital evolution into a GHZ state with a fidelity of 0.55 was demonstrated for a 4-qubit system.

In this letter we demonstrate a versatile superconducting quantum processor featuring
high connectivity with programmable qubit-qubit couplings mediated 
by a bus resonator, and experimentally produce GHZ states with up to 10
qubits using this quantum processor. The
resonator-induced qubit-qubit couplings result in a phase shift that is
quadratically proportional to the total qubit excitation number, evolving
the participating qubits from an initially product state to the GHZ state
after a single collective interaction, irrespective of the number of the
entangled qubits~\cite{Zheng2001}. We characterize the multipartite entanglement by 
quantum state tomography achieved by synchronized local manipulations and detections 
of the entangled qubits, and measure a fidelity of 
$0.668\pm 0.025$ for the 10-qubit GHZ state, which confirms the genuine tenpartite
entanglement~\cite{Guhne2009} with 6.7 standard deviations ($\sigma$).
We also implement parallel entangling operations mediated by the resonator, 
simultaneously generating three Einstein-Podolsky-Rosen (EPR) pairs;
this feature was previously suggested in the
context of ion traps~\cite{Sorensen1999} and quantum dots coupled to an optical cavity~\cite{Imamoglu1999}, 
but experimental demonstrations are still lacking. 

\begin{figure}[t]
\includegraphics[clip=True,width=2.8in]{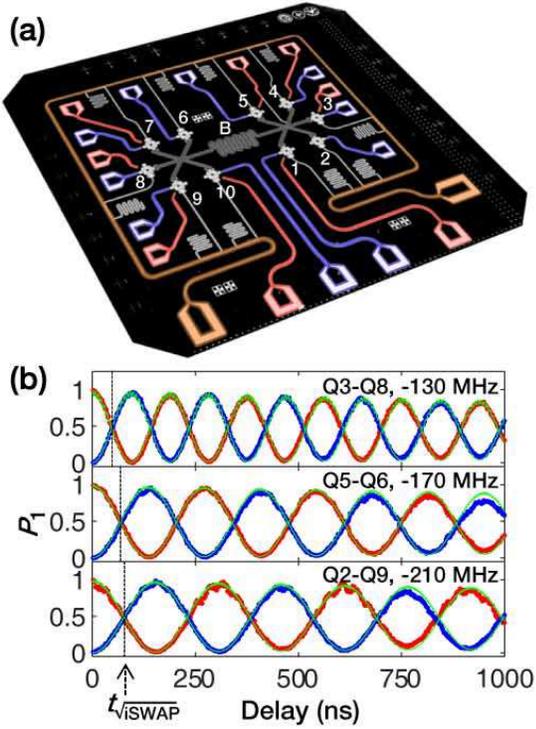}
\caption{
(a) False-color circuit image showing 10 superconducting qubits (star shapes) 
interconnected by a central bus resonator $B$ (grey). 
Each qubit has its own microwave line (red) for XY control and flux bias line (blue) for Z control, 
except for $Q_2$ and $Q_6$, which share the microwave lines of neighboring qubits.
Each qubit has its own readout resonator, which couples to
the circumferential transmission line (orange) for simultaneous readout.
(b) Parallel intra-pair SE interactions for 
$Q_{3}$-$Q_{8}$ (top), $Q_{5}$-$Q_{6}$ (middle), and $Q_{2}$-$Q_{9}$ (bottom)
at the corresponding detunings as indicated. The anti-correlated, 
time-modulated occupation probabilities $P_{10}$ (red dots)
and $P_{01}$ (blue dots) of each pair indicate that energy is
exchanged within the pair~\cite{Ansmann2009}, 
undisturbed by what happen in the other two pairs: 
6 qubits in three pairs are measured simultaneously and 
we ignore outcomes of the other qubits for the two-qubit data shown in each panel. 
All directly measured qubit occupation probabilities are 
corrected for elimination of the measurement errors~\cite{Zheng2017}.
Lines (green) are numerical simulations. 
The small high-frequency oscillations in the simulation curve (green) for $Q_{3}$-$Q_{8}$ 
are due to the relatively small qubit-resonator detuning. 
These small oscillations can be reduced by using a larger detuning, but
at the price of a smaller intra-pair SE interaction strength. 
}
\label{fig1}
\end{figure}

The superconducting quantum processor is illustrated in Fig.~\ref{fig1}(a), which is constructed as
10 transmon qubits ($Q_{j}$ for $j$ = 1 to 10), with resonant frequencies
$\omega_{j}/2\pi$ tunable from 5 to 6 GHz, symmetrically
coupled to a central resonator ($B$), whose resonant frequency is fixed at 
$\omega_B/2\pi \approx $ 5.795 GHz. Measured qubit-resonator ($Q_j$-$B$) coupling strengths
$g_{j}/2\pi$ range from 14 to 20 MHz 
(see Supplemental Material~\cite{supp} for details on device, operation, and readout)~\cite{Lucero2012}.
The central resonator serves as a multipurpose actuator, 
enabling controlled long-range logic operations, scalable multiqubit entanglement, and quantum state transfer.
In the rotating-wave approximation and ignoring the crosstalks between qubits (see Supplemental Material~\cite{supp}), 
the Hamiltonian of the system is given by
\begin{equation}
H /\hbar = \omega_B a^{+}a + \sum_{j=1}^{10}{\left[\omega_{j}|1_j\rangle \langle 1_j|
+ g_{j}(\sigma _{j}^{+}a+\sigma _{j}^{-}a^{+})\right]},  \label{eq1}
\end{equation}
where $\sigma _{j}^{+}$ ($\sigma _{j}^{-}$) is the raising (lowering)
operator of $Q_{j}$ and $a^{+}$ ($a$) is the creation (annihilation)
operator of $B$.

The qubit-qubit coupling can be realized through the superexchange (SE)
interaction~\cite{Trotzky2008} mediated by the bus resonator $B$~\cite{Zheng2000,Osnaghi2001,Majer2007,Dewes2012}.
With multiplexing we can further arrange multiple qubit pairs 
at different frequencies to turn on the intra-pair SE interactions simultaneously.
To illustrate this feature, we consider three qubit pairs, $Q_{k}$-$Q_{k^\prime}$, 
$Q_{l}$-$Q_{l^\prime}$, and $Q_{m}$-$Q_{m^\prime}$,
detuned from resonator $B$ by $\Delta_{j}$ ($\equiv \omega_j - \omega_B$, and $\omega_j = \omega_{j^\prime}$)
for $j = k$, $l$, and $m$, respectively,
while all other qubits are far detuned and can be neglected for now. 
In the dispersive regime and when the resonator $B$ is initially in the
ground state, it will remain so throughout the procedure and the effective Hamiltonian for the qubit pairs is
\begin{multline}
H_{1}/\hbar=\sum_{j \in \{k,l,m\}}{\lambda_j\left( \sigma_{j}^{-}\sigma_{j^\prime}^{+} +\sigma_{j}^{+}\sigma_{j^\prime}^{-}\right)} \\
+ \sum_{j \in \{k,l,m\}}\left [\frac{g_j^2}{\Delta_{j}}\left| 1_j\right\rangle\left\langle 1_j\right| 
+ \frac{g_{j^\prime}^2}{\Delta_{j}}\left| 1_{j^\prime}\right\rangle\left\langle 1_{j^\prime}\right|\right],  \label{eq2}
\end{multline} 
where $\lambda_j = \frac{g_j g_{j^\prime}}{\Delta_j}$, $|\Delta_{j}|\gg g_j, \, g_j^\prime$, 
and $|\Delta_{j_1}-\Delta_{j_2}| \gg  \lambda _{j_1},\, \lambda _{j_1}^\prime, 
\, \lambda _{j_2}, \, \lambda_{j_2}^\prime$ for $j_1,\, j_2 \in \{k, l, m\}$ and $j_1\neq j_2$.
With this setting, the resonator $B$ is simultaneously used for three 
intra-pair SE processes; the inter-pair couplings are effectively switched off due to large detunings between different pairs.

With the fast Z control on each qubit, coupling between any two qubits can be dynamically turned on and off 
by matching (intra-pair) and detuning (inter-pair), respectively, their frequencies,
i.e., we can reconfigure the coupling structure in-situ
without modifying the physical wiring of the circuit.
For example, by arranging $\Delta_k$, $\Delta_l$, and $\Delta_m$ in Eq.~(\ref{eq2}) at three
distinct frequencies, we create three qubit pairs 
($Q_{2}$-$Q_{9}$, $Q_{3}$-$Q_{8}$, and $Q_{5}$-$Q_{6}$) featuring programmable
intra-pair SE interactions with negligible inter-pair crosstalks, 
enabling parallel couplings as demonstrated in Fig.~1(b). 
According to the probability evolutions shown in Fig.~\ref{fig1}(b), 
a characteristic gate time, $t_{\sqrt{\text{iSWAP}}}$, 
for each qubit pair can be be identified.

\begin{figure}[t]
\includegraphics[clip=True,width=3.1in]{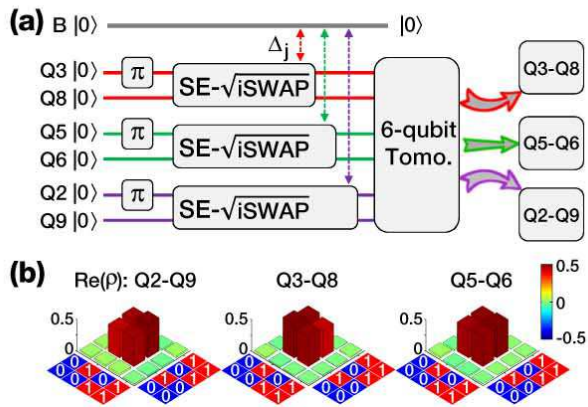}
\caption{
(a) Pulse sequence with detunings listed in Fig.~1(b).
Tomography is performed to reconstruct the 6-qubit density matrix, over which 
we perform partial trace to obtain the reduced density matrix of each EPR pair. 
Each $\pi$-rotation (tomographic $\pi/2$-rotation) pulse has a length of 60 (30) ns 
and a full width half maximum of 30 (15) ns, designed following the
derivative reduction by adiabatic gate (DRAG) control theory~\cite{Motzoi2009}.
(b) Real parts of the reconstructed 2-qubit density matrices
for the three EPR pairs of $Q_{2}$-$Q_{9}$, $Q_{3}$-$Q_{8}$, and $Q_{5}$-$Q_{6}$,
with fidelities (concurrences) of $0.932 \pm {0.13}$ ($0.869 \pm {0.026}$), 
$0.957 \pm {0.010}$ ($0.915 \pm {0.019}$), and $0.951 \pm {0.010}$ ($0.909 \pm {0.019}$), respectively.
For clarity of display, single-qubit $z$-axis rotations are numerically applied to 
$Q_2$ (93$^\circ$), $Q_3$ (165$^\circ$), and $Q_5$ (42$^\circ$) 
to cancel the arguments of the major off-diagonal elements.
}
\label{fig2}
\end{figure}

\begin{figure*}[t]
\includegraphics[clip=True,width=140mm]{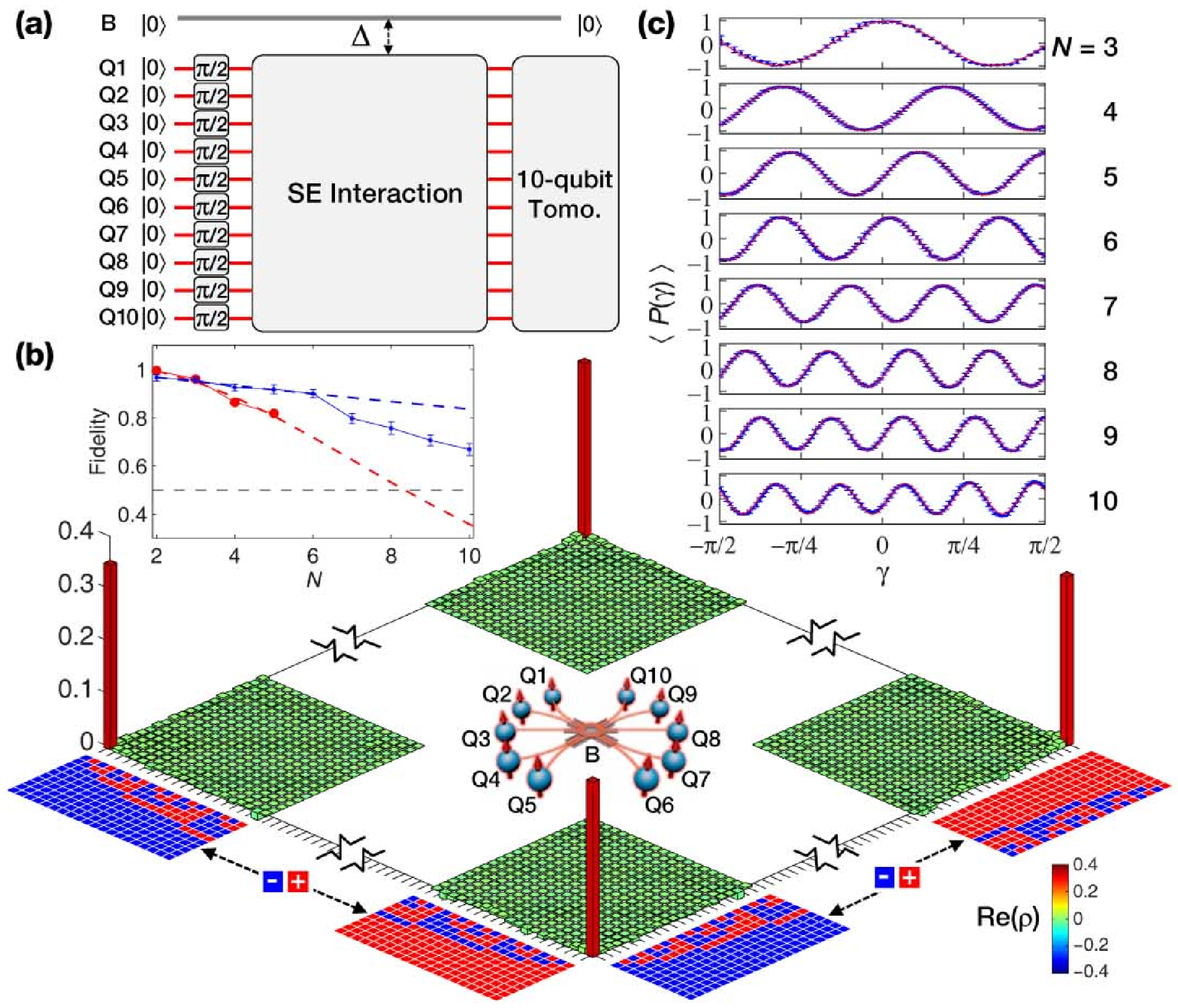}
\caption{
(a) Pulse sequence for the 10-qubit GHZ state with $\Delta/2\pi \approx -140$~MHz.
(b) Partial elements of the measured 10-qubit density matrix, 
with a fidelity of $\textrm{tr}\left(\rho_\textrm{ideal}\rho_\textrm{exp}\right) = 0.668\pm 0.025$ relative to
the ideal GHZ state $|\Psi_1\rangle = \left(|+_1,+_2,...,+_N\rangle +e^{i\varphi}|-_1,-_2,...,-_N\rangle\right)/\sqrt{2}$.
For clarity of display, here a single-qubit rotation around the $x^\prime$ axis by
an angle of $\varphi$ is numerically applied to one of the qubit, 
which cancels the arguments of the dominant off-diagonal elements.
Center inset: Cartoon illustration showing 10 entangled qubits.
Top-left inset: Experimentally measured GHZ fidelity (blue dots) 
and the data adapted from Ref.~\cite{Barends2014} (red dots)
as functions of the qubit number $N$. Error bars are 1$\sigma$. 
Blue and red dashed lines are guides of different error trends.
(c) Parity oscillations observed for the $N$-qubit GHZ states $|\Psi_2\rangle$
defined as superpositions of the basis states $|0_1,0_2,...,0_N\rangle$ and $|1_1,1_2,...,1_N\rangle$ 
with $N = 3$ to 10. The fringe amplitudes are
$0.964\pm0.016$, $0.956\pm 0.018$, $0.935\pm 0.020$, $0.926\pm 0.026$, $0.796\pm 0.023$, 
$0.782\pm 0.025$, $0.729\pm 0.028$ and $0.660\pm 0.032$ from top to bottom.
For $N = 10$, the state preparation and measurement sequence is repeated 81,000
times for a sample size large enough to count the $2^N$ probabilities.
}
\label{fig3}
\end{figure*}

Operating multiple pairs in parallel
naturally produces multiple EPR pairs~\cite{Zheng2000,Osnaghi2001}.
As the pulse sequence shows in Fig.~\ref{fig2}(a), 
three EPR pairs are produced after the completion of all three SE-$\sqrt{\text{iSWAP}}$ gates~\cite{Dewes2012}, 
with the 6-qubit quantum state tomography  
measuring an overall state fidelity of $0.904\pm 0.018$. 
The inferred density matrix $\rho$ is validated by satisfying the physical
constraints of Hermitian, unit trace, and positive semi-definite.
We further perform partial trace on $\rho$ to obtain three 2-qubit reduced density matrices, 
each corresponding to a EPR pair with a fidelity above 0.93 (Fig.~\ref{fig2}(b)).

Remarkably, our architecture allows high-efficiency generation of
multiqubit GHZ states. In contrast to the previous approach
where GHZ states are generated by a series of controlled-NOT (CNOT) gates~\cite{Barends2014}, 
here all the qubits connected to the bus resonator can be entangled 
with a single collective qubit-resonator interaction.
In the theoretical proposal~\cite{Zheng2001,Agarwal1997}, $N$ qubits are assumed to be equally coupled to the resonator
and are detuned from the resonator by the same amount $\Delta$ that is much larger than the qubit-resonator coupling. 
When all qubits are initialized in the same equal superpositions of $0\rangle$
and $|1\rangle$, e.g., $\left(\ket{0}-i\ket{1}\right)/\sqrt{2}$,
the SE interaction does not induce any energy exchange between qubits; instead, it produces a dynamic
phase that nonlinearly depends upon the collective qubit excitation
number $k$ as $k(N+1-k)\theta$, 
where $\theta$ is determined by the effective qubit-qubit coupling strength 
and the interaction time. With the choice $\theta= \pi/2$, this
gives rise to the GHZ state 
$\left(|+_1,+_2,...,+_N\rangle +i|-_1,-_2,...,-_N\rangle\right)/\sqrt{2}$, 
where $|\pm_j\rangle = \left(\ket{0_j} \pm i^{N}\ket{1_j}\right)/\sqrt{2}$~\cite{Zheng2001}.

Here we apply this proposal to our experiment.
We find that, though the qubit-resonator couplings are
not uniform and unwanted crosstalk couplings exist in our circuit, we can optimize
each qubit's detuning and the overall interaction time
to achieve GHZ states with high fidelities as guided by numerical simulation.
The pulse sequence is shown in Fig.~\ref{fig3}(a). We
start with initializing the chosen $N$ qubits in $\left(\ket{0}-i\ket{1}\right)/\sqrt{2}$ by applying $\pi /2$ pulses
at their respective idle frequencies, following which we bias them 
to nearby $\Delta/2\pi \approx -140$~MHz for an optimized duration of approximately twice $t_{\sqrt{\text{iSWAP}}}$. 
The phase of each qubit's XY drive is calibrated according to the rotating frame with respect to $\Delta$,
ensuring that all $N$ qubits are in the same initial state just before their SE interactions are switched on~\cite{Zhong2016,supp}.
After the optimized interaction time, these qubits approximately evolve to the GHZ state 
$|\Psi_1\rangle = \left(|+_1,+_2,...,+_N\rangle + e^{i\varphi} |-_1,-_2,...,-_N\rangle\right)/\sqrt{2}$, 
where 
$\varphi$ may not be equal to $\pi/2$ as in the ideal case with uniform qubit-qubit interactions; 
however, this phase variation does not affect entanglement.
Later on we bias these $N$ qubits back to their idle frequencies;
during the process a dynamical phase $\phi_j$ is accumulated between $|0\rangle$ and $|1\rangle$ of $Q_j$.
Re-defining $|\pm_j\rangle = \left(\ket{0_j} \pm i^{N} e^{i\phi_j} \ket{1_j}\right)/\sqrt{2}$ 
ensures that the above-mentioned formulation of $|\Psi_1\rangle$ remains invariant, which is 
equivalent to a $z$-axis rotation of the $x$-$y$-$z$ reference frame, 
i.e., $x\rightarrow x^\prime$ and $y\rightarrow y^\prime$.
Tracking the new axes is important for characterization of the produced GHZ states. 

Tomography of the produced states requires individually measuring the qubits 
in bases formed by the eigenvectors of the Pauli operators  $X$, $Y$, and $Z$, 
respectively. Measurement in the $Z$ basis can be directly performed.
For each state preparation and measurement event,  
we record the 0 or 1 outcomes of each qubit and do so for $N$ qubits simultaneously;
repeating the state preparation and measurement event thousands of times we count $2^N$ probabilities 
of \{$P_{00...0}$, $P_{00...1}$, ...., $P_{11...1}$\}. Measurement in the 
$X$ ($Y$) basis is achieved by inserting a Pauli $Y$ ($X$) rotation on each qubit before readout. 
All directly measured qubit occupation probabilities are corrected for elimination of the measurement errors~\cite{Zheng2017}.
The $3^N$ tomographic operations and the $2^N$ probabilities for each operation allow
us to 
reconstruct all elements of the density matrix $\rho$ 
(see Supplemental Material~\cite{supp} for various aspects of our tomography technique including measurement stability, 
reliability with reduced sampling size, and pre-processing for minimizing the computational cost).
The resulting 10-qubit GHZ density matrix is partially illustrated in Fig.~3(b), with a fidelity of $0.668\pm 0.025$, 
and the $N$-qubit GHZ fidelity as function of $N$ is plotted in Fig.~3(b) inset. 
The achieved fidelities are well above the threshold for genuine multipartite
entanglement~\cite{Guhne2009}. 

The full tomography technique, though general and accurate, is costly when $N$ is large. 
The produced GHZ states can also be characterized by a shortcut,
since the ideal GHZ density matrix consists of only four non-zero elements in a suitably chosen basis.
To do so, we apply to each qubit a $\pi/2$ rotation around its $y^\prime$ or $x^\prime$ axis, 
transforming $|\Psi_1\rangle$ to 
$|\Psi_2\rangle = \left(|00...0\rangle +e^{i\varphi}|11...1\rangle\right)/\sqrt{2}$ 
(here and below we omit the subscripts of the qubit index 
for clarity). The diagonal elements $\rho_{00...0}$ and $\rho_{11...1}$
can be directly measured; the off-diagonal elements $\rho_{00...0,\,11...1}$ 
and $\rho_{11...1,\,00...0}$ can be obtained by 
measuring the system parity, defined as the expectation value of the operator 
$P(\gamma)=\otimes_{j=1}^{N} (\cos{\gamma}Y^\prime_j+\sin{\gamma}X^\prime_j)$, which is given by
$\langle P(\gamma) \rangle = 2\left|\rho_{00...0,\, 11...1}\right|\cos(N\gamma+\varphi)$ 
for $|\Psi_2\rangle$~\cite{Monz2011}. 
Polarization along the axis defined by $\cos\gamma Y^\prime+\sin⁡\gamma X^\prime$ can be 
measured after applying to each qubit a rotation by an angle $\gamma$ around the $z^\prime$ axis~\cite{Zhong2016}.
The oscillation patterns of the measured parity as functions of $\gamma$ 
confirm the existence of coherence between the states $|00...0\rangle$ and $|11...1\rangle$ (Fig.~3(c)).
The fidelity of the $N$-qubit GHZ state $|\Psi_2\rangle$ can be estimated using the four non-zero elements,
which is $0.660\pm0.020$ for $N=10$. This value agrees 
with that of the GHZ state $|\Psi_1\rangle$ obtained by full state tomography. 

A key advantage of the present protocol for generating GHZ states is its high scalability as demonstrated in Fig.~\ref{fig3}(b).
If limited by decoherence, the achieved fidelity based on the 
sequential-CNOT approach, $F_{\text{N,C}}$, scales approximately as
$F_{\text{N,C}}\propto F_{2,C}^{N^{2}/2}$ at large $N$ (see the red dashed line in Fig.~\ref{fig3}(b) inset), while that based on 
our protocol scales as $F_{\text{N}}\propto F_{2}^{N}$ (blue dashed line).
Here $F_{2,C}$ ($F_{2}$) is quoted as the decoherence-limited fidelity of the CNOT gate (present protocol) involving two qubits.
The falling of the experimental data (blue dots) below the scaling line when $N \ge 6$ is due
to the inhomogeneity of $g_j$ and the crosstalk couplings. 
One can see that, even with the two-qubit gate fidelity above 0.99 
as demonstrated in two recent experiments~\cite{Barends2014, Sheldon2016}, the
coherence performance of the devices does not allow generation of 
10-qubit GHZ state with fidelity above the genuine entanglement threshold
using the sequential-CNOT approach.
 
In summary, our experiment demonstrates the viability of the multiqubit-resonator-bus
architecture 
with essential functions including
high-efficiency entanglement generation and parallel logic operations. 
We deterministically generate the 10-qubit GHZ state,
the largest multiqubit entanglement ever created in solid-state systems, which is
verified by quantum state tomography for the first time as well.
In addition, our approach allows instant \textit{in situ} rewiring of the qubits,
featuring all-to-all connectivity that is critical in a recent proposal~\cite{Li2017}. These unique features
show the great potential of the demonstrated approach for scalable quantum information processing.\\

This work was supported by the National Basic
Research Program of China (Grants No. 2014CB921201
and No. 2014CB921401), the National Natural
Science Foundations of China (Grants No. 11434008,
No. 11374054, No. 11574380, No. 11374344, and
No. 11404386), the Fundamental Research Funds for the
Central Universities of China (Grant No. 2016XZZX002-01),
and the National Key Research and Development Program of
China (Grant No. 2016YFA0301802). S. H. was supported by
National Science Foundation (Grant No. PHY-1314861).
Devices were made at the Nanofabrication Facilities at
Institute of Physics in Beijing, University of Science and
Technology of China in Hefei, and National Center for
Nanoscience and Technology in Beijing.

\clearpage
\renewcommand\thefigure{S\arabic{figure}}
\renewcommand\theequation{S\arabic{equation}}
\renewcommand\thetable{S\arabic{table}}

\setcounter{figure}{0}
\setcounter{equation}{0}
\setcounter{table}{0}

\renewcommand\thefigure{S\arabic{figure}}
\renewcommand\theequation{S\arabic{equation}}
\renewcommand\thetable{S\arabic{table}}
\renewcommand\thesection{\arabic{section}}
\renewcommand\thesubsection{\thesection.\arabic{subsection}}

\begin{center}
{\noindent {\bf Supplementary Material for\\
``10-qubit entanglement and parallel logic operations
with a superconducting circuit''}}
\end{center}

\section{1. General information of the experimental setup}

\begin{figure*}[t]
  \centering
  \includegraphics[width=6.5in,clip=True]{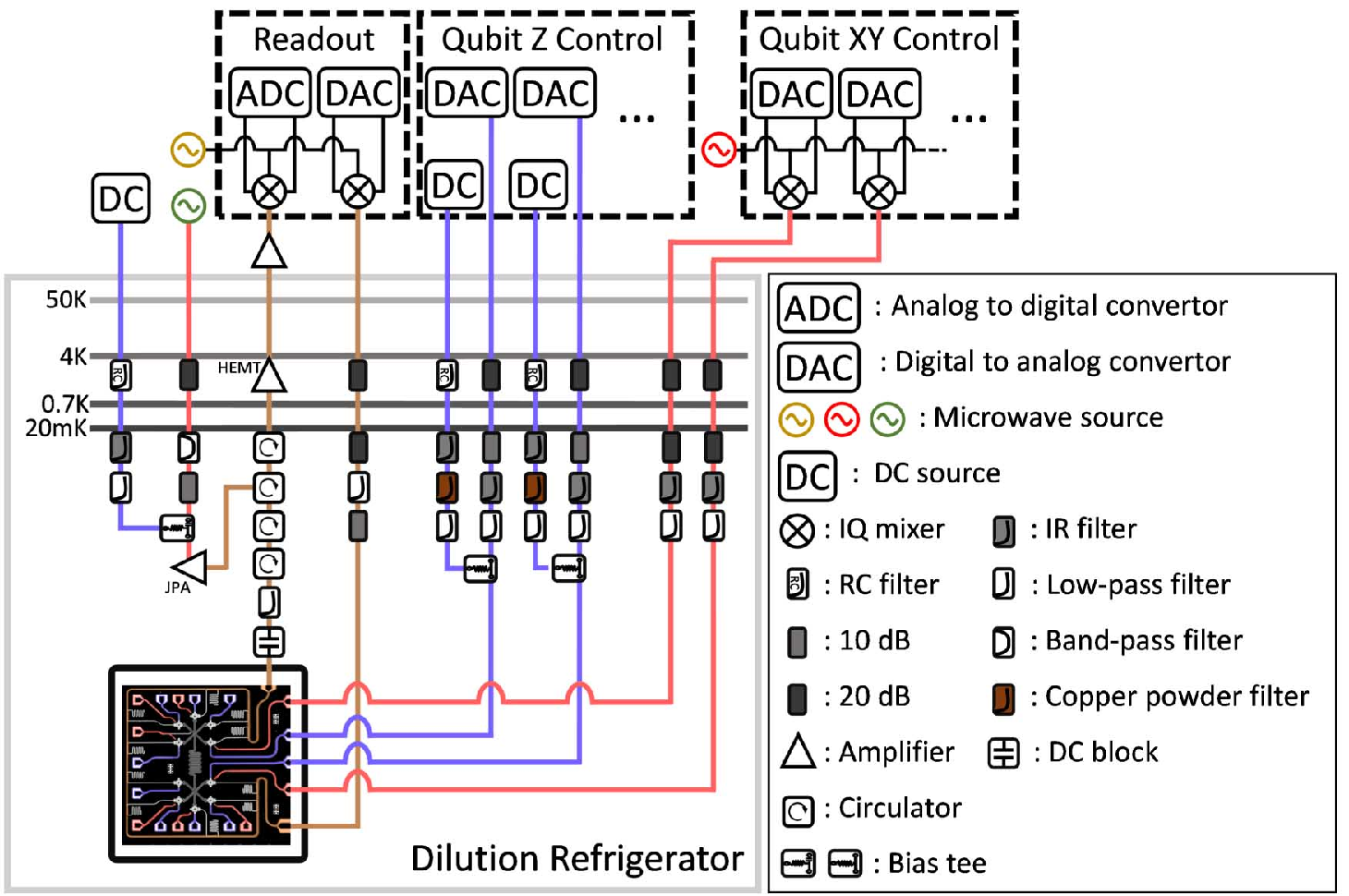}\\
  \caption{\textbf{Overview of the experimental setup.}
	Diagram detailing all the control electronics, wiring, and filtering for the experimental setup. 
	On the top from left to right, we show electronics for the Josephson parametric amplifier (JPA) control, qubit readout, qubit Z control, 
	and qubit XY control, respectively. The 
	DAC in the readout box and the nearby microwave source
	output a ten-tone microwave pulse targeting all qubits' readout resonators. 
	The readout signal is amplified sequentially by the JPA, high electron mobility transistor (HEMT), 
	and room temperature amplifiers before being demodulated by the analog digital converter (ADC).  
	In the qubit Z control box, each DC source outputs a static frequency offset and each DAC offers the dynamic frequency tuning of a qubit. 
	DACs in the qubit XY control box and the nearby microwave source together output microwaves for the XY control of qubits. 
	All control lines go through various stages of attenuation and filtering to prevent unwanted noises from disturbing the operation of the device.
	}
	\label{fig.wiring}
\end{figure*}

Figure~\ref{fig.wiring} provides an overview of the experimental setup, with 
individual components being addressed below in detail.
In this section we also discuss some characteristics of 
the microwave and Z bias crosstalks of our device, 
and introduce the numerical simulation which takes into account the device characteristics.

\subsection{1.1. Device}
\begin{table*}[b]
\centering
\caption{Qubit characteristics.
$\omega_j^{0}$ is the maximum resonant frequency of $Q_j$, i.e., 
the resonant frequency at the qubit's sweetpoint where dephasing is minimum.
$\omega_I$ is the interaction frequency for qubits to entangle, 
which is about $2\pi \times 140$~MHz below the resonator frequency $\omega_B$.
Qubit coherence parameters including the lifetime $T_{1,j}$, the Ramsey Gaussian dephasing time $T_{2,j}^\ast$,
and the spin echo Gaussian dephasing time $T_{2,j}^\textrm{SE}$ of $Q_j$ are measured at $\omega_I$ where other qubits are far detuned.
$\omega_j$ is the idle frequency of $Q_j$, where the $X/2$ gate for state preparation and
the tomographic pulses for qubit measurement are applied.
$g_j$ is the coupling strength between $Q_j$ and the resonator $B$, defined in the interaction
Hamiltonian $\hbar g_j (\sigma_j^+ + \sigma_j^-)(a^\dagger + a)$, where $\sigma_j^+$ ($\sigma_j^-$)
is the raising (lowering) operator of $Q_j$ and $a^\dagger$ and $a$ are field operators of resonator $B$.
Fidelity of the $X/2$ gate on $Q_j$ is characterized by RB 
at its idle frequency $\omega_j$; we also simultaneously run RBs on 8 qubits to characterize
the $X/2$ gate fidelities for these 8 qubits. 
Each qubit is measured through its own readout resonator that connects to a common circumferential transmission line, and
$\omega_j^{r}$ is the resonant frequency of $Q_j$'s readout resonator $R_j$, with
the $Q_j$-$R_j$ coupling strength noted as $g_j^r$. 
$\delta \omega_j^{m}$ is the AC stark shift of $Q_j$ during its measurement, i.e., the qubit frequency
shifts downwards from $\omega_j$ as its readout resonator $R_j$ becomes populated.
The readout pulse is 1~$\mu$s-long, and carries 10 tones targeting the 10 readout resonators.
$n_j^{r}$ ($= |\delta \omega_j^{m}/2\chi_j|$, where $\chi_j$ is the dispersive frequency shift of the
readout resonator when qubit changes state from $|0\rangle$ to $|1\rangle$) is the maximum average photon number in $Q_j$'s 
readout resonator $R_j$ as excited by the readout pulse for qubit measurement, which decays
at the end of the readout pulse with a leakage rate of $\kappa_j^{r}$ for 
$R_j$ returning to the vacuum state and being ready for the next cycle.
$F_{0,j}$ ($F_{1,j}$) is the probability of correctly reading out $Q_j$ in $|0\rangle$ ($|1\rangle$) 
when it is prepared in $|0\rangle$ ($|1\rangle$), which are used to correct the measured qubit probabilities~\cite{Zheng2017}. 
$Q_3$'s readout fidelities apply to the entangling experiments with $N \leq 9$.
We use two concatenated $X/2$ gates to prepare the qubit in $|1\rangle$.
Values in parenthesis, if available, are one standard deviation.}
\begin{ruledtabular}
\begin{tabular}{ c c c c c c c c c c c c c }
                          && $Q_1$    & $Q_2$    & $Q_3$    & $Q_4$    & $Q_5$    & $Q_6$ & $Q_7$ & $Q_8$ & $Q_9$ & $Q_{10}$ \\
\hline
$\omega_j^{0}/2\pi$ (GHz) && $5.782$ & $5.831$ & $5.828$ & $5.780$ & $5.760$ & $5.863$ & $5.780$ & 6.004 & 5.893 & 5.930 \\
$\omega_I/2\pi$ (GHz)     && \multicolumn{10}{c}{$\approx 5.655$} \\
$T_{1,j}$ ($\mu$s)            && 27.2 & 24.4 & 10.9 & 15.0 & 19.2 & 23.7 & 13.8 & 11.8 & 17.1 & 22.0 \\
$T_{2,j}^\ast$ ($\mu$s)       && 2.9  & 2.8  & 2.8  & 2.2  & 2.6  & 1.8  & 1.1  & 2.1  & 1.7  & 4.4  \\
$T_{2,j}^\textrm{SE}$ ($\mu$s)&& 11.8 & 10.6  & 10.0 & 10.8 & 11.7 & 8.9  & 8.0  & 8.0  & 7.9  & 11.8 \\
$\omega_j/2\pi$ (GHz)     && 5.080 & 5.467 & 5.657 & 5.042 & 5.179 & 5.605 & 4.960 & 5.260 & 5.146 & 5.560 \\
$g_j/2\pi$ (MHz)          && 14.2  & 20.5  & 19.9  & 20.2  & 15.2  & 19.9 & 19.6 & 18.9 & 19.8 & 16.3 \\
\hline
$X/2$       fidelity      && 0.9985(2) & 0.9992(1) & 0.9984(1) & 0.9987(2) & 0.9991(1) & 0.9964(5) & 0.9987(1) & 0.9980(3) & 0.9988(3) & 0.9989(1) \\
Simultaneous              &&  &  &  &  &  &  &  &  &  &  \\
$X/2$ fidelity            && {0.9978(7)} & {0.9980(2)} & {0.9953(5)} & {0.9955(5)} & {0.9985(2)} & {0.9962(3)} & {--}        & {--}        & {0.9979(2)} & {0.9931(12)} \\
\hline
$\omega_j^{r}/2\pi$ (GHz) && 6.509 & 6.541 & 6.615 & 6.614 & 6.635 & 6.694 & 6.691 & 6.794 & 6.809 &6.891 \\
$g_j^r/2\pi$ (MHz)        && 41.3  & 39.9  & 40.6  & 38.2  & 38.5  & 40.4 & 41.8 & 40.9 & 40.2 & 38.7 \\
$\delta\omega_j^{m}/2\pi$ (MHz) && 31.1 & 32.7 & 21.1 & 46.5 & 9.0 & 45.1 & 22.5 & 19.5 & 26.0 & 70.2 \\
$n_j^{r}$                 && 92    & 59    & 31    & 180   & 30    & 81 & 93 & 74 & 103 & 203 \\
$1/\kappa_j^{r}$ (ns)     && 291   & 275   & 272   & 348   & 223   & 284 & 248 & 266 & 299 & 242 \\ 
$F_{0,j}$                 && 0.921 & 0.955 & 0.982 & 0.974 & 0.962 & 0.988 & 0.950 & 0.970 & 0.961 & 0.971 &\\
$F_{1,j}$                 && 0.867 & 0.915 & 0.904 & 0.928 & 0.927 & 0.917 & 0.922 & 0.880 & 0.894 & 0.934 &\\
\end{tabular}
\end{ruledtabular}
\label{tab1}
\end{table*}

\begin{figure*}[t]
  \centering
  \includegraphics[width=7in,clip=True]{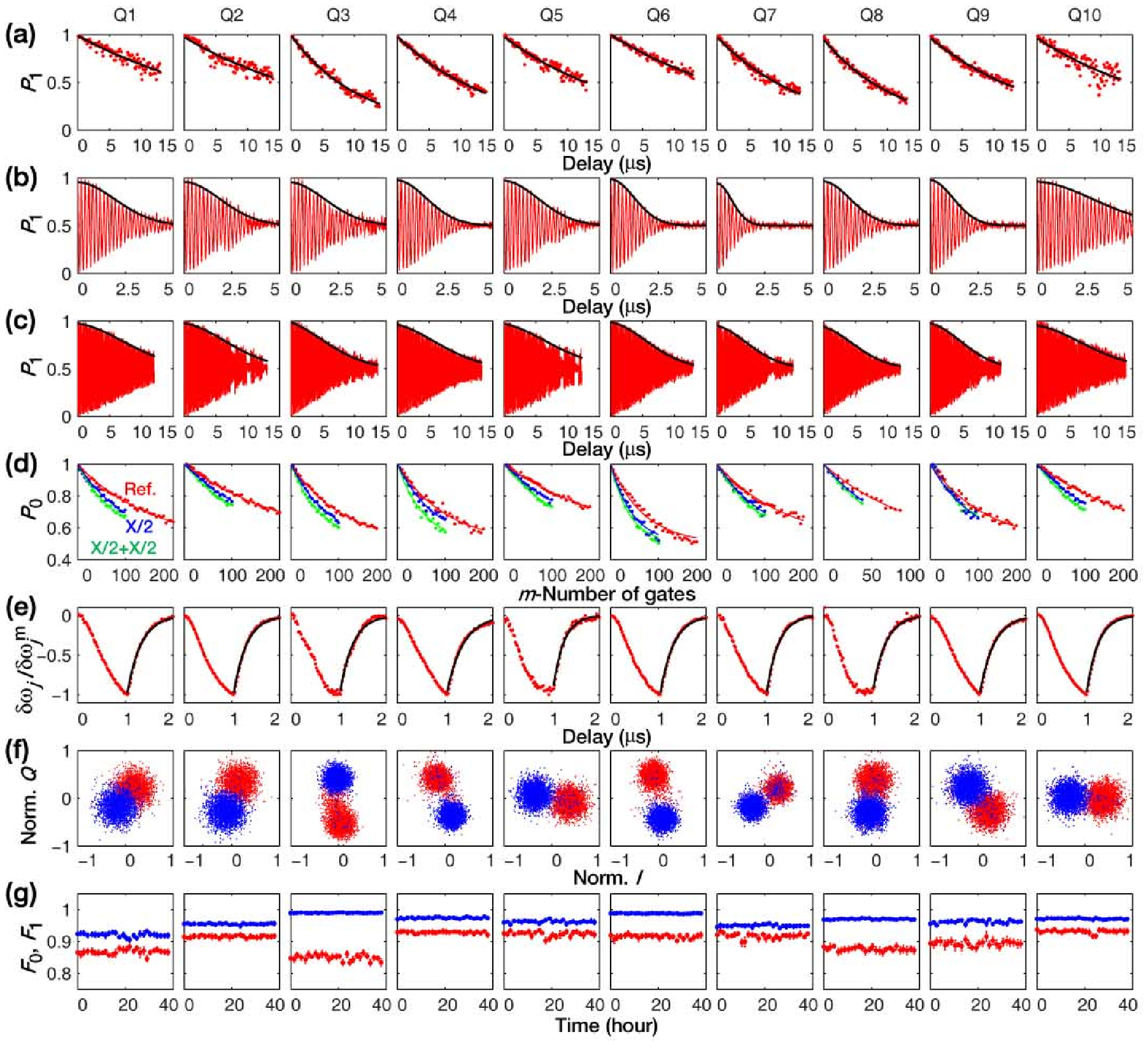}\\
  \caption{\textbf{Qubit characteristics.}
	(a) Qubit energy decay measurement (red dots) with the corrected $|1\rangle$-state probability, $P_1$, 
	as a function of the decay time, $t$.
	Black lines are fits according to $P_{1,j} \propto \exp{\left(-t/T_{1,j}\right)}$ for $Q_j$.
	(b) Qubit Ramsey interference measurement (red dots), where the delay time $t$ is between two $\pi/2$ pulses.
	We fit the envelope of $Q_j$'s Ramsey fringe with 
	$P_{1,j}^\textrm{Env} - 0.5 \propto \exp{\left[-t/2T_{1,j}-\left(t/T_{2,j}^\ast\right)^2\right]}$.
	(c) Qubit spin echo measurement (red dots), where an additional $\pi$ pulse is inserted in the middle 
	of the delay time $t$ compared with the Ramsey sequence. We fit the envelope of $Q_j$'s interference fringe with 
	$P_{1,j}^\textrm{Env} - 0.5 \propto \exp{\left[-t/2T_{1,j}-\left(t/T_{2,j}^\textrm{SE}\right)^2\right]}$.
	(d) RB characterizing the $X/2$ and $X/2+X/2$ gates. Shown are the $|0\rangle$-state probabilities 
	averaged over $k = 20$ sequences for the reference (red), the $X/2$ gates interleaved with the reference (blue), and 
	the $X/2+X/2$ gates interleaved with the reference (green)
	as functions of the $m$-Number of gates. Probability values drop exponentially with $m$
  due to the randomized accumulation of gate-specific errors. Dots are experimental data and lines are fits.
	(e) The normalized AC stark shift during the qubit-state measurement (see $\delta\omega_j^{m}$'s value in Tab.~\ref{tab1}).
	Two consecutive readout pulses are used in this experiment following procedures described elsewhere~\cite{Jeffrey2014}.
	Black lines are exponential fits giving the photon leakage rates $\kappa_j^{r}$.
	(f) The shifted and normalized $I$-$Q$ data for single-shot qubit-state differentiation, where blue (red) dots are
	measured $I$-$Q$ values when qubit is prepared in $|0\rangle$ ($|1\rangle$). There are 3000 dots (repetitions) for each color (qubit state).
	$Q_3$'s $I$-$Q$ data apply to the entangling experiments with $N \le 9$.
	(g) Qubit $|0\rangle$- and $|1\rangle$-state readout fidelities with error bars, $F_{0,j}$ (blue) 
	and $F_{1,j}$ (red), monitored over a 40-hour period. 
	$Q_3$'s readout fidelities apply to the entangling experiment with $N = 10$ as $Q_4$ is simultaneously measured (see below).
	We note that $F_{1,j}$ is usually very stable over a period
	of a few hours, but could vary more than displayed due
	to random movements of two-level defects and fluctuations of other environmental factors, 
	all of which affect qubit parameters and degrade the gate fidelity for the $|1\rangle$-state generation.
	As such we have to constantly monitor the system performance and repeat tune-up procedures if necessary.
	}
	\label{fig.perf}
\end{figure*}

Our sample is a superconducting circuit consisting of 10 Xmon qubits interconnected by
a bus resonator, fabricated with two steps of aluminum deposition 
as described elsewhere~\cite{Song2017}.
There is no crossover layer to connect segments of grounding planes on the circuit chip.
Instead, aluminum bonding wires are manually applied as many as possible to
reduce the impact of parasitic slotline modes. 
The bus resonator $B$ is a half-wavelength coplanar waveguide resonator with
a resonant frequency fixed at $\omega_B /2\pi \approx 5.795$~GHz, which is measured with all coupling qubits 
staying in $|0\rangle$ at their respective idle frequencies (see below).
The resonator has 10 side arms, each is capacitively coupled to an Xmon qubit 
with the coupling strength $g_j$ listed in Tab.~\ref{tab1}.
The Xmon qubit is a variant of the transmon qubit, each with an individual
flux line for dynamically tuning its frequency and a microwave drive ($Q_2$, $Q_4$, and $Q_6$ 
share other qubits' microwave lines in this experiment) for controllably 
exciting its $|0\rangle \leftrightarrow |1\rangle$ transition. 
The Xmon qubit reaches its maximum resonant frequency $\omega_j^0$ at the sweetpoint, where it is insensitive
to flux noise and exhibits the longest phase coherence time.
$\omega_j^0$ for $j = 1$ to 10 in our device are around or slightly above $\omega_B$,
whose values are roughly estimated through the flux-biased spectroscopy measurement.
In this experiment all qubits are initialized to the ground
state $|0\rangle$ at their respective idle frequencies $\omega_j/2\pi$ that 
spread in the range from 4.96 to 5.66 GHz, corresponding to 840 to 140~MHz below $\omega_B/2\pi$ (Tab.~\ref{tab1}),
where single-qubit rotations and the qubit-state measurement are performed. 
For initialization we simply idle all qubits for more than 200 $\mu$s~\cite{comment,Johnson2012,Riste2012,Jin2015,Chen2016};
For the entangling operation we dynamically bias all target qubits from
their idle frequencies $\omega_j$ to the interaction frequency $\omega_I$ 
for a specified interaction period,
following which we bias all these qubits back to their $\omega_j$ for measurement.
Qubit coherence performance at $\omega_I$ can be found in Tab.~\ref{tab1} and in Fig.~\ref{fig.perf}.

\subsection{1.2. XY control}
Our instrument has 7 (expendable to more) independent XY signal channels controlled by digital analog converters (DACs): 
3 channels are selected to output two tones
per channel and the rest 4 channels are programmed to output a single tone per channel,
for a total of XY controls with 10 tones targeting 10 qubits. 
The 10 tones are generated with 10 sidebands mixed with a continuous microwave 
whose carry frequency is $\omega_c/2\pi = 5.324$~GHz.
Microwave leakage is minimized with this standard mixing method as calibrated by the room temperature electronics.

In our setup, $Q_1$ and $Q_2$ share $Q_1$'s on-chip XY line connecting to the 1st two-tone XY channel, 
$Q_3$ and $Q_4$ share $Q_3$'s on-chip XY line connecting to the 2nd two-tone XY channel, 
and $Q_6$ and $Q_7$ share $Q_7$'s on-chip XY line connecting to the 3rd two-tone XY channel.
For the two-tone XY signals that are supposed to simultaneously act on the two qubits, 
we sequentially place two rotation pulses, each with a single tone
targeting one of the two qubits. Each $\pi$-rotation ($\pi/2$-rotation) pulse has 
a length of 60 (30) ns and a full width half maximum of 30 (15) ns, designed following the
derivative reduction by adiabatic gate (DRAG) control theory~\cite{Motzoi2009}. 
We perform randomized benchmark (RB) on the two qubits simultaneously, and verify that
overlapping in time the two rotation pulses through the two-tone XY channel 
is not a problem except for $Q_6$ and $Q_7$, which might be due to the large sideband used for $Q_7$.

\begin{figure*}[t]
  \centering
  \includegraphics[width=7in,clip=True]{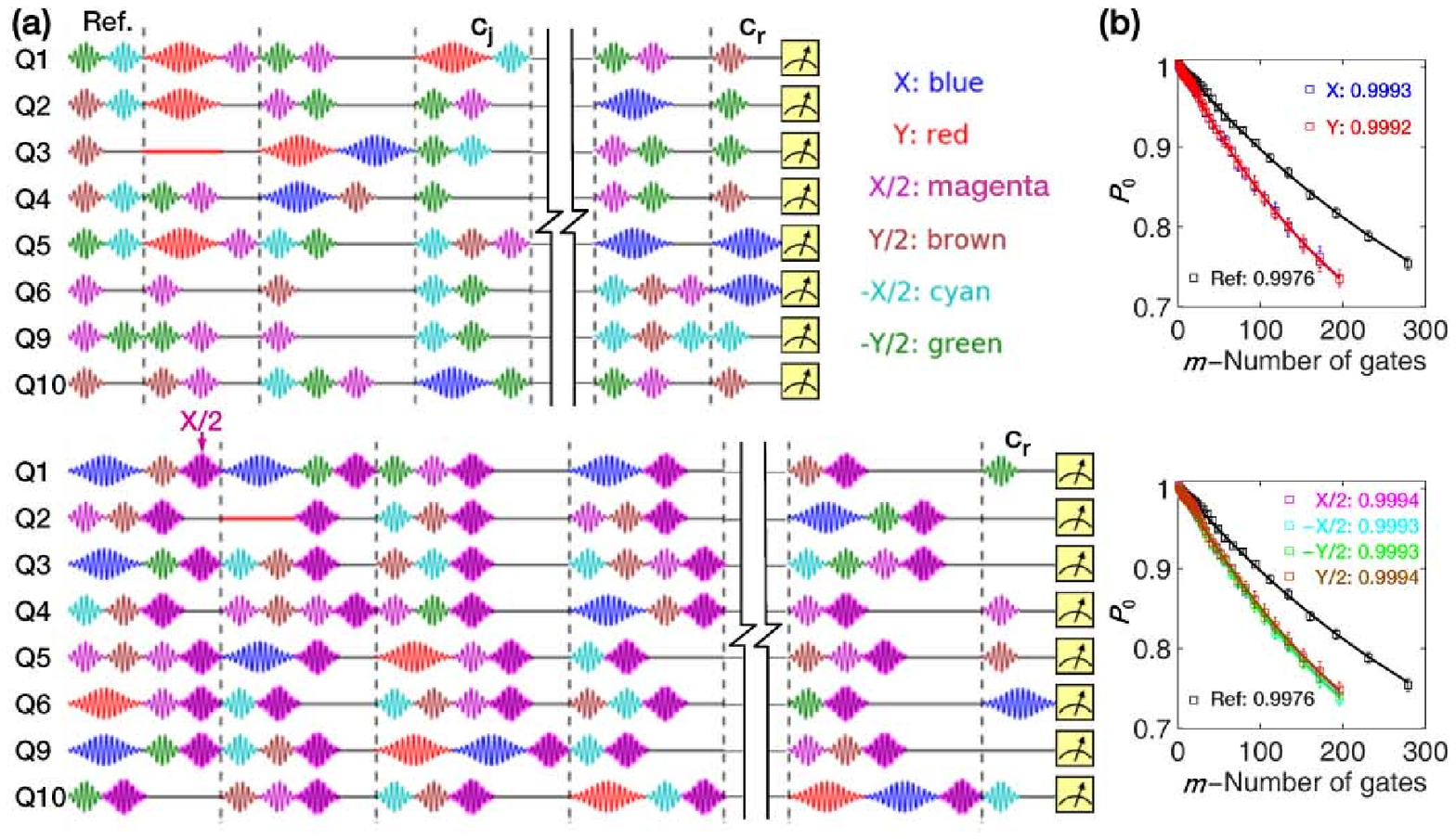}\\
  \caption{\textbf{Single-qubit RBs.}
	(a) Pulse sequences for single-qubit RBs running on 8 qubits simultaneously.
	Top: One representative reference gate sequence. Each sequence has up to $m$
	single-qubit Cliffords $C_j$ (divided by vertical dashed lines) 
	and the idling gates, the latter of which are implemented to ensure that 
	the $j$-th Cliffords on all qubits take the same time slot. 
	Single-qubit rotation gates (sinusoids) used to assemble $C_j$ are color-coded on the right 
	(the $I$ gate is represented by a horizontal red line).
	The final $C_r$ gate returns each qubit to $|0\rangle$.
	At the end of the pulse sequence we measure all 8 qubits simultaneously for
	$2^8$ occupation probabilities, based on which we perform partial trace over other qubits' indices
	for one piece of data on the target qubit; we sum over all data of $k = 20$ random pulse sequences
	for an exponential fit.
	Bottom: One representative gate sequence with the $X/2$ gates (bold magenta sinusoids) inserted.
	On average, each single-qubit Clifford $C_j$ consists of 
	0.375 gates from $X$, $Y$, and $I$, 1.5 gates from $\pm X/2$ and $\pm Y/2$, and 
	approximately 1 idling gate to ensure synchronization among qubits.
	Pulse lengths of the $X$, $Y$, and $I$ gates are 60 ns; lengths of the $\pm X/2$ and $\pm Y/2$ gates are 30 ns; 
	length of the idling gate for synchronization is 30 ns. In total the average 
	duration of a single-qubit Clifford is about 98 ns.
  (b) Single-qubit RBs for $Q_5$ ($k = 30$), where we shorten the pulse length to 20 ns, set $\alpha = 0$, and
	optimize the detuning to be $-0.37$~MHz. All single-qubit gate fidelities are above 0.999.}
	\label{fig.RB}
\end{figure*}

The idle frequencies $\omega_j$ of the 10 qubits are detuned from each other to minimize
the microwave crosstalk during single-qubit rotations. 
For two qubits sharing the same XY line, the cross-resonance interaction 
reported elsewhere~\cite{Chow2011} is not a major factor due to the large detuning.  
For each qubit, we optimize the quadrature correction term with DRAG coefficient $\alpha$  
to minimize leakage to higher levels~\cite{Kelly2014}, yielding the $X/2$ ($\pi/2$ rotation around the $x$ axis)
gate fidelities no less than 0.998 for all 10 qubits as verified 
by RB on each qubit (Tab.~\ref{tab1} and in Fig.~\ref{fig.perf}).
We also select eight qubits and simultaneously perform RBs on them (see pulse sequences in Fig.~\ref{fig.RB}(a)),
finding that the $X/2$ gate fidelities remain reasonably high, no less than 0.993
($Q_7$ is not selected due to the above-mentioned large sideband issue; 
$Q_8$'s DAC control has a smaller memory, with a maximum sequence length only half of the others').
We also perform RB simultaneously with shorter pulse sequences (smaller $m$-Number of gates) on $Q_8$ and $Q_9$,
two neighboring qubits with the idle frequencies being very close, 
and find that the $X/2$ gate fidelities remain no less than 0.997.

We note that further optimization of the single-qubit gates are possible 
by shortening the gate time and introducing a slight detuning to the XY pulse to minimize the phase error~\cite{Kelly2014}. 
We carry out the optimization procedure on $Q_5$ and obtain fidelities of 
the typical single-qubit gates, $X$, $Y$, $\pm X/2$, and $\pm Y/2$, all above 0.999 (Fig.~\ref{fig.RB}(b)). 

We note that although single-qubit gates are optimized at the qubits'
idle frequencies $\omega_j$, the gate performance may slightly degrade after a big square pulse used to 
tune the qubit frequency due to the finite rise-up time 
on the order of a few to a few tens of nanoseconds for an ideally sharp step-edge. 
In our experiment, after biasing the qubits
from the interaction frequency $\omega_I$ back to their idle frequencies $\omega_j$, 
we wait for another 10 ns before applying single-qubit gates. 
We also use the GHZ tomography data to benchmark the gate fidelities of our $\pi/2$ rotations.
For $N=7$, the density matrix of $|\Psi_1\rangle$ with four major elements in the $|\pm\rangle$ basis
has a fidelity of $0.796\pm0.021$. After applying a $\pi/2$ rotation to each qubit we transform $|\Psi_1\rangle$ to
$|\Psi_2\rangle$ in the $|0\rangle$ and $|1\rangle$ basis, whose density
matrix is measured to exhibit a fidelity of $0.771\pm0.022$ (each 7-qubit full tomography is done within 40 minutes). Therefore we estimate
that our $\pi/2$ rotation after the big square pulse for entanglement has an average fidelity around $(0.771/0.796)^{1/7} > 0.995$,
while a more detailed numerical simulation suggests that the average $\pi/2$ gate fidelity is $> 0.993$.

\subsection{1.3. Z control}
Our instrument has 10 (expendable to more) independent Z signal channels controlled by DACs, 
which give us the full capability of simultaneously 
tuning all 10 qubits' resonant frequencies.
To correct for the finite rise-up time of
an ideally sharp edge of a square pulse,
we generate a step-edge output from the DAC and
capture the waveform with a high-speed sampling oscilloscope.
The measured response of the step-edge gives two time constants 
describing the response of the room-temperature wirings, 
based on which we use de-convolution to correct for desired step-edge pulses.
Imperfection due to the cryogenic wirings are partially compensated 
using the qubit's transition frequency response influenced 
by a step-edge Z bias as a caliber~\cite{Hofheinz2009}.

The effective Z bias of a qubit due to a unitary bias applied to other qubits' Z lines is calibrated, 
which yields the Z-crosstalk matrix 
$\tilde{M}_Z$ as
\begin{equation}
\label{eq1}
\left[
{\tiny
\begin{smallmatrix}
 1     & -0.023 & -0.001 &  0.008 &  0.028 &  0.020 &  0.009 &  0.011 &  0.003 & -0.008  \\
-0.018 &  1     &  0.050 &  0.024 &  0.008 &  0.003 &  0.000 & -0.002 &  0.005 & -0.016  \\
-0.021 & -0.081 &  1     &  0.054 &  0.022 &  0.012 &  0.003 &  0.001 & -0.004 & -0.017  \\
 0.017 &  0.055 &  0.080 &  1     & -0.024 & -0.012 & -0.003 &  0     &  0.003 &  0.013  \\
 0.016 &  0.019 &  0.009 & -0.003 &  1     & -0.015 & -0.002 &  0.005 &  0.009 &  0.014  \\
 0.001 & -0.002 & -0.004 & -0.005 & -0.025 &  1     &  0.009 &  0.022 &  0.013 &  0.004  \\
 0.001 &  0     & -0.003 & -0.006 & -0.028 & -0.046 &  1     &  0.078 &  0.035 &  0.008  \\
-0.004 &  0     &  0.001 &  0.003 &  0.013 &  0.020 &  0.025 &  1     & -0.029 & -0.006  \\
-0.012 & -0.006 & -0.002 &  0     &  0.008 &  0.015 &  0.016 &  0.065 &  1     & -0.014  \\
 0.002 &  0.011 &  0.015 &  0.015 &  0.029 &  0.023 &  0.010 &  0.008 & -0.011 &  1
\end{smallmatrix}
}
\right].
\end{equation}
With the Z biases applied to the 10 qubits 
written in a column format as $\tilde{Z}_\textrm{applied}$
and the actual Z biases sensed by these 10 qubits
written as $\tilde{Z}_\textrm{actual}$, we have the mapping relation of $\tilde{Z}_\textrm{actual} = \tilde{M}_Z \cdot\tilde{Z}_\textrm{applied}$.
The Z crosstalks reach maximum at about 8\% between two neighboring qubits. 
We note that the Z crosstalks may not contribute to the GHZ state errors
as we iteratively fine-tune the Z bias of each qubit within a small range for optimal GHZ entanglement.

\subsection{1.4. Qubit readout}
\begin{figure*}[t]
  \centering
  \includegraphics[width=6in,clip=True]{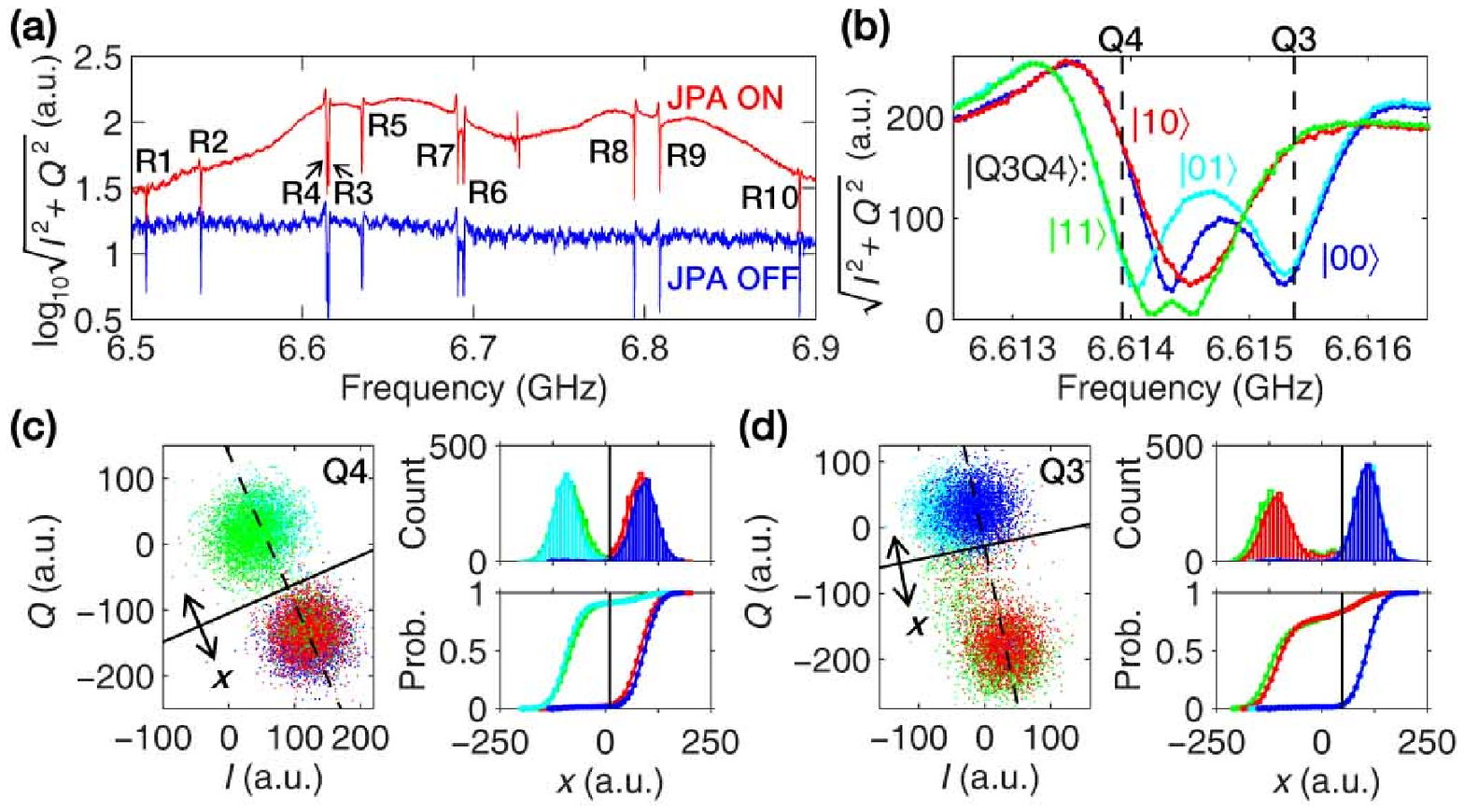}\\
  \caption{\textbf{Multiqubit readout.}
	(a) Signal transmission spectra while all qubits are in $|0\rangle$. Shown are the amplitudes of
	the demodulated signal as functions of signal frequency when the JPA
	is ``ON'' (red) and ``OFF'' (blue). All readout resonators, labeled
	from $R_1$ to $R_{10}$, are visible as dips on the spectra ($R_3$ and $R_4$ are very close). The vertical
	difference between the red and blue lines manifests the JPA's capability of amplifying signals.
	(b) Detailed signal transmission spectra when $Q_3$ and $Q_4$ are prepared
	in one of the four computational states of $|00\rangle$ (blue), $|01\rangle$ (cyan), $|10\rangle$ (red), and $|11\rangle$ (green).
	It is possible to differentiate both qubits' state if the readout tones and powers are properly chosen as indicated by the vertical dash lines.
	(c) Left panel: Demodulated $I$-$Q$ data of $Q_4$ when $Q_3$ and $Q_4$ are prepared in
	one of the four computational states (color-coded as in \textbf{b}). 
	Top-right panel: Distribution histogram along the $x$ axis as defined by the dash line in the left panel.
	The vertical line indicates the separatrix to differentiate $|0\rangle$ and $|1\rangle$. 
	Bottom-right panel: Measurement visibility by integrating the histogram data along the $x$ axis.
	(d) Readout of $Q_3$ when $Q_3$ and $Q_4$ are prepared in
	one of the four computational states (color-coded as in \textbf{b}).
	The readout crosstalk is visible in the left panel of \textbf{d}.
	But the relative movement of $Q_3$'s $|0\rangle$-state blobs depending on $Q_4$ in $|0\rangle$ (blue blob)
	and in $|1\rangle$ (cyan blob) are almost vertical to the $x$ axis, and so are $Q_3$'s $|1\rangle$-state
	blobs. Therefore $Q_4$'s effect on $Q_3$ is negligible with our choice of
	the solid dividing line for state differentiation.
	The readout tone used for $Q_3$ here is slightly different from that used when $Q_4$ is not measured.
	With this choice $Q_3$'s readout fidelity slightly decreases, but we have 
	verified that with the readout correction we can
	prepare various product states of $Q_3$, $Q_4$, and $Q_5$ with high-fidelity single-qubit
	gates, and perform one- and two-qubit state tomography, with the fidelities
	of all reconstructed density matrices being around or above 0.99.}
	\label{fig.readout}
\end{figure*}

Besides the above-mentioned 7 XY signal channels for the qubit control, 
our instrument has an XY signal channel that can output a
readout pulse with multiple tones achieved by sideband mixing;
this readout pulse is captured by a room-temperature 
ADC, which simultaneously demodulates the multiple (up to 20) tones and returns a pair of $I$ and $Q$ values for each tone.
An impedance-transformed JPA operating at 20~mK is used
before the ADC to enhance the signal-to-noise ratio. 
The signal-line impedance of the JPA continuously varies, 
in a manner of the Klopfenstein taper, so that
the environmental characteristic impedance changes from 50 to 15 $\Omega$,
which gives a JPA bandwidth of more than 200~MHz centered around 6.72~GHz. 
The fabrication procedure can be found elsewhere~\cite{Song2017}.
The JPA can be switched ``ON'' and ``OFF'' by turning on and off, respectively, an appropriate
pump tone that is about twice the signal frequency.
The signal transmission spectra
with the JPA in the states ``ON'' (red line) and ``OFF'' (blue line) are displayed in Fig.~\ref{fig.readout},
where the 10 dips correspond to the 10 readout resonators.
The amplification band of the JPA, identified by the vertical difference between
the red and blue lines, is tunable with a DC bias applied to the JPA. 

The readout pulse is 1 $\mu$s-long, with the input tones and the power at each tone optimized for high-fidelity readout. 
The $j$-th tone of the readout pulse, where $j$ is up to 10 in this experiment, populates $Q_j$'s readout resonator $R_j$ with 
an average photon number of $n_j^r$ in 1 $\mu$s, which dispersively
interacts with $Q_j$ with the coupling strength $g_j^r$ and shifts $Q_j$'s frequency
downwards by an amount of $\delta\omega_j^{m}$. 
Reversely, the qubit state affects the state of its readout resonator, which is 
encoded in the $I$-$Q$ values at the tone $j$ of the transmitted readout pulse.
At the end of 1 $\mu$s, photons in $Q_j$'s readout resonator
leak into the circumferential transmission line at the
rate of $\kappa_j^{r}$, and the readout resonator returns to the ground state before the next sequence cycle starts.

Repeated readout signals amplified by the JPA are demodulated at room temperature, yielding 
the $I$-$Q$ points at each tone on the complex plane
forming two blobs to differentiate the states $|0\rangle$ and $|1\rangle$ of each qubit 
(see Tab.~\ref{tab1} and in Fig.~\ref{fig.perf}). 
The probabilities of correctly reading out each qubit
in $|0\rangle$ and $|1\rangle$ are listed in Tab.~\ref{tab1}.

We note that the readout resonators of $Q_3$ and $Q_4$
are very close in frequency, and so are those of $Q_6$ and $Q_7$. We carefully
choose the readout tones and powers to minimize the readout crosstalk if $Q_3$ and $Q_4$ are both being measured,
which has a slight side-effect that the readout visibility of $Q_3$ drops a little bit compared with the case when $Q_4$ is not being measured (Fig.~\ref{fig.readout}).
Nevertheless, our readout choice for minimizing the crosstalk is fully verified by preparing various product states 
of $Q_3$, $Q_4$, and $Q_5$ with high-fidelity single-qubit gates and performing single- and two-qubit state tomography, 
with the fidelities of all reconstructed density matrices being around or above 0.99.
	
\subsection{1.5. $XX$-type crosstalk coupling}
Due to insufficient crossover bonding wires to tie the ground segments on-chip,
we experience unwanted microwave crosstalk coupling between nearest-neighbor qubits. 
The crosstalk coupling is calibrated by measuring the qubit-qubit energy swap process
around the interaction frequency $\omega_I$.

\begin{figure}[t]
  \centering
  \includegraphics[width=3.5in,clip=True]{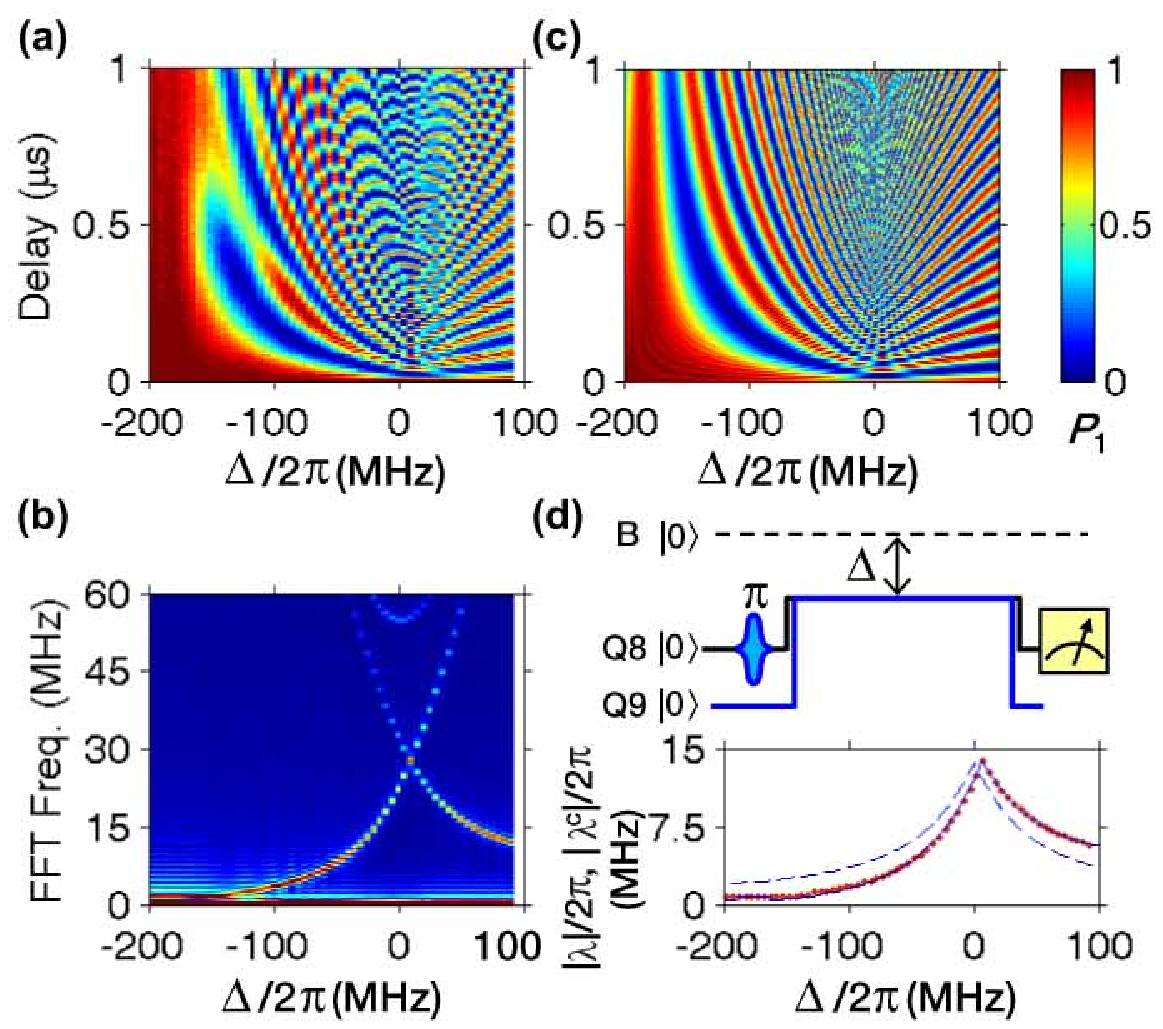}\\
  \caption{\textbf{$Q_8$-$Q_9$'s energy-swap dynamics showing the
	qubit-qubit interaction including the contribution from the neighboring $XX$-type crosstalk.} After
	pumping $Q_8$ to $\ket{1}$, we detune $Q_8$ and $Q_9$ both to $\Delta$ for
	a time of delay. $Q_8$ is measured afterwards. 
	(a) Measured $P_{1}$ as a function of both $\Delta$ ($x$ axis) and delay ($y$ axis). 
	(b) Fourier transform of the data in \textbf{a} along the $y$ axis. 
	(c) Numerical simulation of the measurement in \textbf{a} taking into account the $XX$-type crosstalk interaction.
	(d) Top: Pulse sequence for the measurement in \textbf{a}. 
	Bottom: Amplitude of the total interaction (dots) as a function
	of $\Delta$ obtained from the lower branch in \textbf{b}. 
	For comparison, the interaction from the Fourier transform of the numerical data in \textbf{c} (solid line) and the theoretical SE strength $\lambda$
	(dashed line) are also illustrated. The SE interaction changes sign across $\Delta = 0$, so that
	the net qubit-qubit interaction amplitude is antisymmetric.}\label{fig.xtalk}
\end{figure}

To understand the crosstalk coupling, we measure in detail
the energy swap process of $Q_8$ and $Q_9$ as a function of the qubit detuning from the resonator $\omega_B$, 
with the result shown in Fig.~\ref{fig.xtalk}(a) ($Q_8$ and $Q_9$ are chosen since they
are the nearest-neighbor qubits with two highest sweetpoint frequencies $\omega_8^0$ and $\omega_9^0$):
We excite $Q_8$ to $\ket{1}$ and then detune both $Q_8$ and $Q_9$ simultaneously to the same 
detuning $\Delta$ from the resonator, with $\Delta$ being varied;
by subsequently monitoring the $|1\rangle$-state population of $Q_8$, $P_1$, as a function of the interaction time,
we obtain the energy-swap dynamics of the system at various detunings.
The qubit-qubit interaction strength can be inferred from the oscillation period of $P_1$.

We find that, in addition to the resonator mediated SE coupling $\lambda$ (see Eq.~(2) of the main text) which changes sign across $\omega_B$, 
a direct $XX$-type coupling with a magnitude of $\approx 2\pi \times 2.1$~MHz, 
named as $\lambda_{8,9}^c$ for $Q_8$-$Q_9$, must be taken into account to explain our experimental data. 
Figure~\ref{fig.xtalk}(b) shows the Fourier transform of
the data in Fig.~\ref{fig.xtalk}(a) along the $y$ axis,
based on which the net qubit-qubit interaction  
as a function of $\Delta$ can be inferred, as shown in Fig.~\ref{fig.xtalk}(d).
Overall the experimental data agree well with the numerical 
simulation taking into account the extra $XX$-type coupling $\lambda^c$ (Figs.~\ref{fig.xtalk}(c) and (d)). 
The $XX$-type crosstalk couplings $\lambda_{j,j^\prime}^c$ between other nearest-neighbor qubits ($Q_j$ and $Q_{j^\prime}$) are roughly
estimated as the differences between the measured qubit-qubit coupling strengths and theoretical
SE interaction strengths, as given in Tab.~\ref{tab2}.

\begin{table}
\centering
\caption{Nearest-neighbor $XX$-type crosstalk couplings.}
\begin{ruledtabular}
\begin{tabular}{ c c c c c c c c c c c c c }
&& $Q_1$-$Q_2$  & $Q_2$-$Q_3$  & $Q_3$-$Q_4$  & $Q_4$-$Q_5$  & $Q_5$-$Q_6$ \\
$\lambda_{j,j^\prime}^c/2\pi$ (MHz) && 1.7 & 2.6 & 2.3 & 2.2 & 0.2 \\
\hline
&& $Q_6$-$Q_7$ & $Q_7$-$Q_8$ & $Q_8$-$Q_9$ & $Q_9$-$Q_{10}$ & $Q_{10}$-$Q_1$ \\
$\lambda_{j,j^\prime}^c/2\pi$ (MHz) && 2.2 & 2.3 & 2.1 & 2.1 & 0.06 \\
\end{tabular}
\end{ruledtabular}
\label{tab2}
\end{table}

\subsection{1.6. Numerical simulation}

The numerical simulation in Fig.~\ref{fig.xtalk}(c) is based on 
the Hamiltonian shown in Eq.~(1) of the main text with the
additional nearest-neighbor crosstalk coupling term 
$\sum_{j,j^\prime}{\lambda_{j,j^\prime}^c\left(\sigma_{j}^{-}\sigma_{j^\prime}^{+} +\sigma_{j}^{+}\sigma_{j^\prime}^{-}\right)}$.
The decoherence impact, if considered, is included using the Lindblad master equation
taking into account a Markovian environment to avoid numerical complexity.
Two characteristic decay times, the energy relaxation
time $T_{1,j}$ and the pure dephasing time $T_{\varphi,j}$, are used
for qubit $j$. However, the non-Markovian $1 / f$ character of the phase noise
prevents us from directly using the $T_{2,j}^\ast$ values listed in Tab.~\ref{tab1}. 
Meanwhile, $T_{2,j}^\ast$s are measured when the qubits are detuned 
and uncoupled, where each qubit frequency depends on the flux in its own transmon loop.
On the opposite, during the resonant coupling as done in Fig.~\ref{fig.xtalk}, the two qubits of the pair 
form a new system with eigenenergies that depend very weakly on each qubit flux. In other words, 
the eigenenergies at an anticrossing are always flat and are much less sensitive to noise, 
and the two hybridized levels form a kind of decoherence-free subspace~\cite{VION2017}.
Consequently, we use empirical $T_{\varphi,j}$ values ($\approx 10 T_{2,j}^\ast$) to capture the decoherence impact if necessary,
which ensures a good agreement between the numerical results and the experimental data.

\begin{figure}[t]
  \centering
  \includegraphics[width=2in,clip=True]{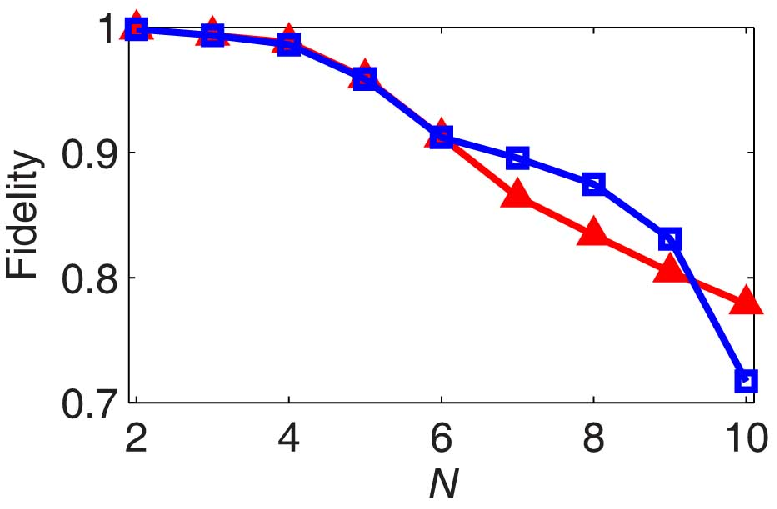}\\
  \caption{\textbf{GHZ state fidelity by numerical simulation} 
	versus the qubit number $N$ with (red) 
	and without (blue) including the XX-type nearest-neighbor couplings as listed in Tab.~\ref{tab2}. 
	The simulation does not consider decoherence,
	and only covers the multiqubit dynamical process for entanglement. 
	The initial condition of the simulation is that each qubit is at $\omega_I$ 
	(140~MHz below the resonator frequency) and is in the state of $(|0\rangle-i|1\rangle)/\sqrt{2}$.
	We check the multiqubit density matrix by ignoring the resonator's outcome right after the optimal interaction time.
	Other relevant parameters needed for the simulation can be found in Tab~\ref{tab1}.  
	If considering decoherence based on the Lindblad master equation using $T_{\varphi,j} \approx 10 T_{2,j}^\ast$,
	the simulation gives a GHZ fidelity of $\approx 0.819$ for $N=7$ 
	while the experimental result	is 
	$0.796\pm0.021$.}\label{fig.sim}
\end{figure}

During the numerical optimization of the parameters for generating GHZ states,
effects of the $XX$-type nearest-neighbor couplings as listed in Tab.~\ref{tab2} are investigated. 
It is found that the introduction of the crosstalk couplings 
lowers the GHZ fidelities for $N > 6$ but actually raises the fidelity for $N = 10$ (Fig.~\ref{fig.sim}).

\section{2. GHZ entanglement generation}

\subsection{2.1. Multiqubit GHZ phase calibration at $\omega_I$}
For the multiqubit GHZ entanglement, it is critical that the phase of each qubit's XY
drive is calibrated according to the rotating frame at $\omega_I$, after taking into account the
extra dynamical phases accumulated during the frequency adjustment of all qubits.
Here we follow the approach as done previously~\cite{Zhong2016}: For simplicity we consider the product state of two qubits as
$\frac{1}{\sqrt{2}}\left(|0\rangle + |1\rangle\right) \otimes \frac{1}{\sqrt{2}}\left(|0\rangle + e^{i\varphi}|1\rangle\right)$,
where the extra $\varphi$ on the second qubit is seen right after the two qubits are placed on-resonance at $\omega_I$;
we intend to find a way to adjust $\varphi$ to be zero.
With the interaction Hamiltonian as $\hbar \lambda (|01\rangle \langle10| + |10\rangle \langle01|)$ (see Eq.~(2) of the main text),
the amplitudes of $|01\rangle$ and $|10\rangle$
then oscillate in time, as described (in the rotating frame) by the unitary transformation  
$U_\textrm{int} = \left[\begin{smallmatrix}
1 & 0 & 0 & 0 \\
0 & \cos(\lambda t) & -i\sin(\lambda t) & 0 \\
0 & -i\sin(\lambda t) & \cos(\lambda t) & 0 \\
0 & 0 & 0 & 1 \\
\end{smallmatrix}\right]$.
At $t = \pi/ 4|\lambda|$, where $\lambda$ is negative, the two-qubit state evolves to
$|00\rangle/2 + (e^{i\varphi}+i)/(2\sqrt{2}) |01\rangle + (i e^{i\varphi} + 1)/(2\sqrt{2}) |10\rangle + e^{i\phi}|11\rangle/2$,
which gives equal probabilities for the four two-qubit computational states with $\varphi = 0$.

\begin{figure}[t]
  \centering
  \includegraphics[width=3.2in,clip=True]{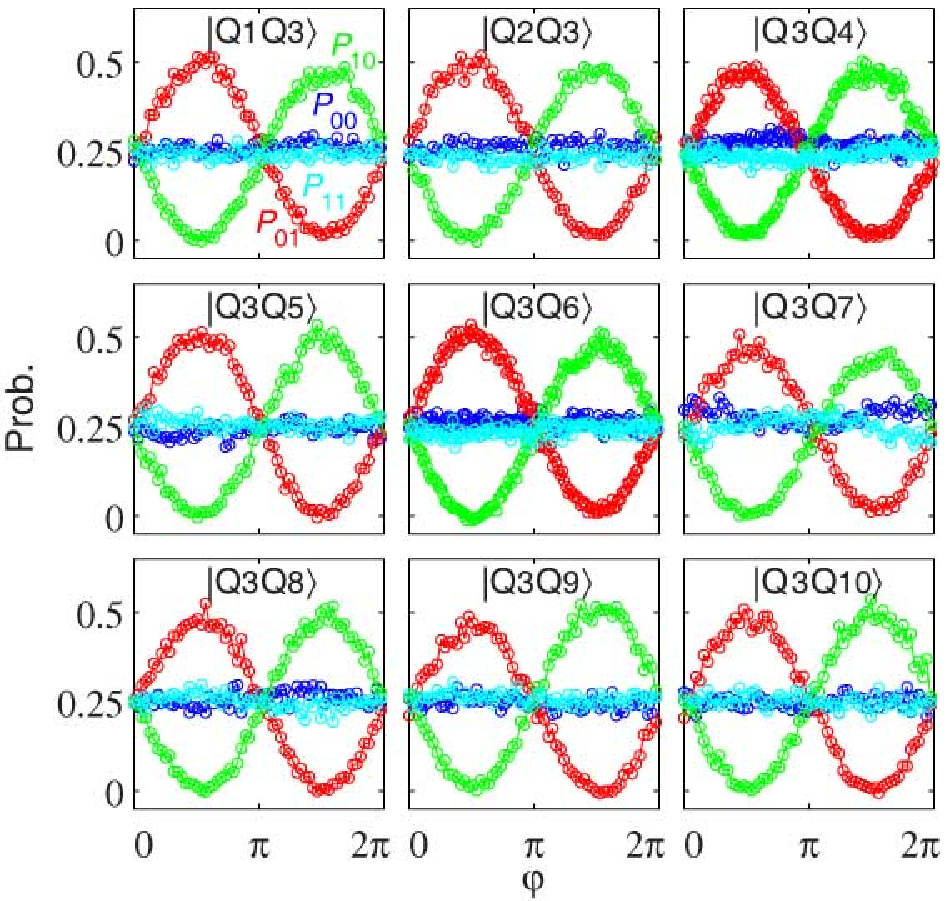}\\
  \caption{\textbf{Pairwise phase calibrations with $Q_3$ as the reference.}
	After driving both qubits with the $\pi/2$ microwave pulses, we tune them to $\omega_I$ for an interaction time of $t = \pi/ 4|\lambda|$, 
	where $\lambda$ is the net qubit-qubit interaction at $\omega_I$. Afterwards the two-qubit occupation probabilities are measured.
	For the two $\pi/2$ microwave pulses, we vary the phase $\varphi$ of the second qubit's microwave for the data shown in each panel with the qubit order as displayed.
	The occupation probabilities, $P_{00}$ (blue), $P_{01}$ (red), $P_{10}$ (green), and $P_{11}$ (cyan), as functions of $\varphi$ are 
	shown after the phases of all qubits' microwaves referenced to $Q_3$'s are corrected.}\label{fig.phase}
\end{figure}

Experimentally we choose $Q_3$ as the reference and adjust the phase of the other qubit's microwave; 
we perform the check pairwisely, with the data after all phase calibrations shown in Fig.~\ref{fig.phase}.

\subsection{2.2. Pulse optimization}

Experimentally we scan over each qubit's detuning and the overall interaction time 
to optimize the GHZ state fidelity. Varying these parameters and checking the resulting GHZ
density matrix is the most straightforward way, but performing the multiqubit tomography is time consuming when 
the qubit number becomes large. Alternatively we only repeatedly measure the reduced density matrix on 
a selected number of qubits, according which we optimize the pulse parameters. 

\begin{figure}[t]
  \centering
  \includegraphics[width=2in,clip=True]{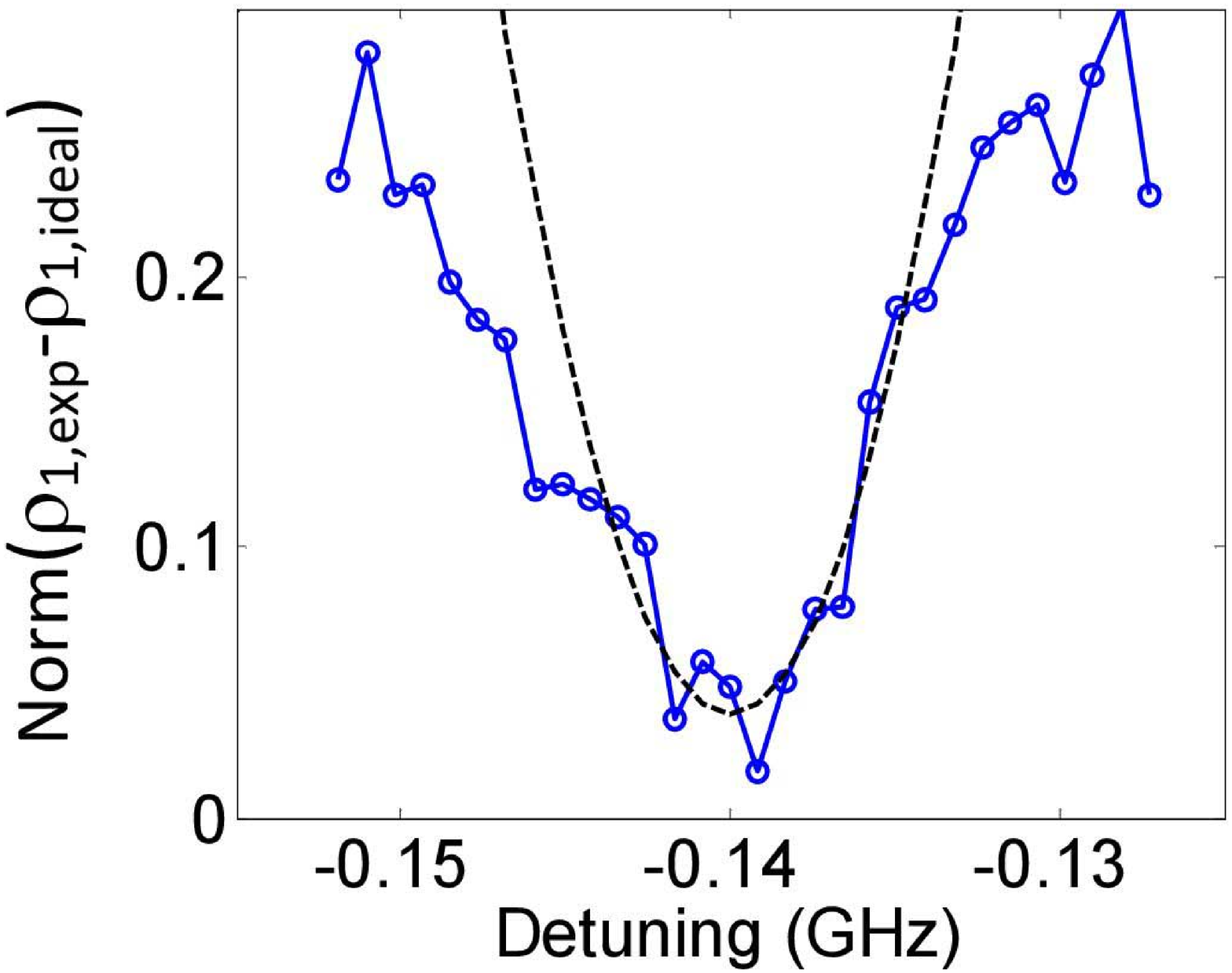}\\
  \caption{\textbf{Optimizing $Q_1$'s detuning in the 10-qubit GHZ experiment.}
	After preparing the GHZ state, we only measure $Q_1$'s density matrix $\rho_{1,\textrm{exp}}$ 
	and compare it with $\rho_{1,\textrm{ideal}}$ by 
	calculating the norm of the difference of the two matrices (circles). We find the best 
	detuning value (x-axis) by locating the minimum of the fitted curve (dashed line).}
	\label{fig.Q1OPT}
\end{figure}

For example, in order to optimize $Q_1$'s detuning, 
we perform the single-qubit tomography on $Q_1$ while ignoring outcomes of all other qubits, 
which, in the ideal case, would yield the one-qubit density matrix as $\rho_{1,\textrm{ideal}} = 
\left(\begin{smallmatrix} 0.5&0\\0&0.5\protect \end{smallmatrix}\right)$.
We compare the experimental matrix with $\rho_{1,\textrm{ideal}}$, and find 
the best detuning at the minimum of the norm of the difference 
of the two matrices, as shown in Fig.~\ref{fig.Q1OPT}. 
Other qubits' parameters are optimized in a similar way.

\section{3. 10-qubit tomography}

All directly measured qubit occupation probabilities are 
corrected for elimination of the measurement errors 
before any further processing~\cite{Zheng2017}. 
With all the corrected probability data we perform unconstrained linear inversion 
to obtain an initial guess of the density matrix, 
which might be unphysical, e.g., with small negative eigenvalues. 
Then we extract a Hermitian, unit trace, and positive semi-definite density matrix 
that is closest in distance to this initial guess~\cite{Wang2011}.
We note that the state fidelity of the inferred matrix 
with respect to the ideal GHZ matrix 
typically drops by 0.03$\sim$0.04 (in absolute value) 
after it is validated.

\subsection{3.1. Effect of reduced sample size}
The $N$-qubit tomography takes $3^N$ tomographic operations, and for each operation 
$2^N$ occupation probabilities of the $N$-qubit computational states are measured. 
Once a GHZ state is generated, a tomographic operation is appended, following which
the single-shot measurement yields a binary outcome for each qubit and for $N$ qubits simultaneously; 
running the whole pulse sequence, which includes the state preparation, the tomographic operation, and the measurement, 
once is one sampling event. Therefore 
a sufficiently large sample size, i.e., repeating the same pulse sequence many times
for many synchronized binary outcomes of all $N$ qubits, is necessary to precisely count all $2^N$ probabilities,
which would significantly slow down the measurement.
For the multiqubit tomography, we maintain a fixed sample size of 3000, which is only about 3 times
$2^N$ when $N = 10$; even in this case the full tomography measurement takes about 40 hours if uninterrupted,
and we have to constantly monitor our measurement to ensure that the system performance is reasonably stable (Fig.~\ref{fig.perf}). 
We note that for $N=10$ a huge number ($3^{10}$) of operations are involved, resulting in a set of over-constrained
equations to solve for the system's density matrix, which may overcome the shortage of an insufficient sample size.

\begin{figure}[t]
  \centering
  \includegraphics[width=2.8in,clip=True]{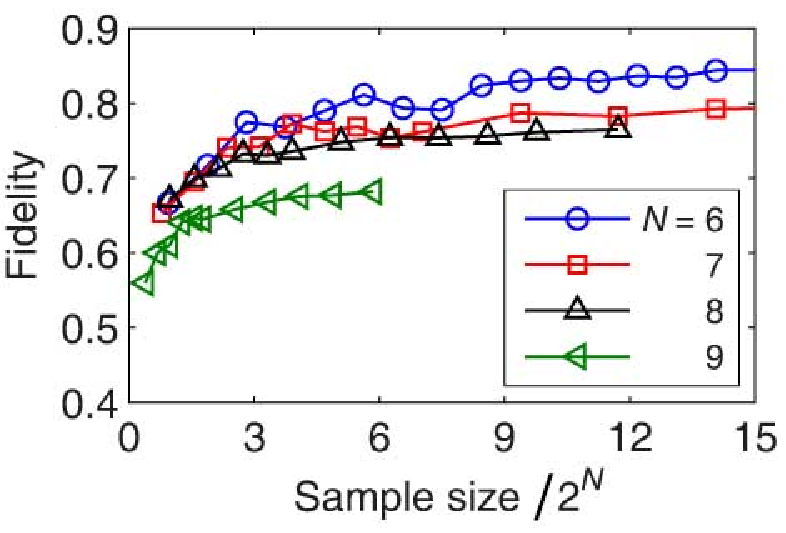}\\
  \caption{\textbf{Tomography with reduced sample size.} 
	Fidelities of reconstructed density matrices as functions of sample size for $N = 6$ to 9.
	The data points for different sample sizes on the same line with a fixed $N$ value are sampled from
	one measurement done with a fixed sample size of 3000: A queue of 3000 simultaneous binary outcomes of $N$ qubits
	are returned during the measurement, based on which we randomly pick up the desired number (a specified sample size) of outcomes from the queue
	and calculate the probabilities; the probabilities are then used to infer the density matrix at the specified sample size.
	During this experiment the state-generation pulse sequences may not be tuned to the optimal.}\label{fig.sampleSize}
\end{figure}

Furthermore, we carry out a test by performing the $N$-qubit tomography with a variable sample size for $N = 6$ to 9.
The results show that, with the sample size around $3\times2^N$, the reconstructed density matrix has a fidelity very close to
that with a sufficiently large sample size (Fig.~\ref{fig.sampleSize}).

\subsection{3.2. Reducing the computation complexity}
Here we use the unitary matrix $U^j$ to describe the $j$-th tomographic operation on the $N$-qubit system whose density matrix is $\rho$,
where both $U^j$ and $\rho$ are of size $2^N \times 2^N$.
The measured probability of the $k$-th computational state after the tomographic operation is therefore
\begin{equation}
\label{eq.Pk}
P_k^j = \langle k | U^j \rho \left(U^j\right)^\dagger |k\rangle = \sum_{l,m=1}^{2^N}{U_{kl}^j \left(U_{km}^j\right)^\ast \rho_{lm}}, 
\end{equation}
where $k = 1$ to $2^N$ indexes the $N$-qubit computational states.
Vectorization of the matrix $\rho$ by stacking its columns into a single column vector $\tilde{\rho}$, 
we have $\tilde{P}^j = \tilde{U}^j \tilde{\rho}$, where $\tilde{P}^j$ is the vector format of $\{P_1^j,\,P_2^j,\,\ldots,\,P_N^j\}$ and
$\tilde{U}^j$ is a $2^N \times 4^N$ matrix replacing the summation terms of $U_{kl}^j$ and $(U_{km}^j)^\ast$ in Eq.~(\ref{eq.Pk}).
Stacking all tomographic operation matrices $\tilde{U^j}$ and all measured probabilities 
$\tilde{P}^j$ for $j = 1$ to $3^N$ into $\tilde{U}$ and $\tilde{P}$, respectively,
we obtain the linear equations of $\tilde{U} \tilde{\rho} = \tilde{P}$, which is used to solve for $\tilde{\rho}$ given $\tilde{P}$ and $\tilde{U}$.

$\tilde{U}$ is a column full rank matrix of size $6^N \times 4^N$. When $N$ approaches 10, 
it becomes extremely difficult to fully load $\tilde{U}$ into a computer's memory and solve for $\tilde{\rho}$. 
Fortunately, only a small fraction of $\tilde{U}$'s elements are
non-zero, so that we can use sparse matrix for storage and employ the pseudo-inverse method. 
With $\tilde{U}^\dagger$ as the Hermitian conjugate of $\tilde{U}$, we have 
$\tilde{U}^\dagger \tilde{U} \tilde{\rho} = \tilde{U}^\dagger \tilde{P}$, 
where $\tilde{U}^\dagger \tilde{U}$ is a symmetric and positive definite matrix of size $4^N \times 4^N$. 
$\tilde{U}^\dagger \tilde{U}$ is not only smaller in size, but also more sparse than $\tilde{U}$, 
which greatly reduce the complexity when solving the equations.

Here we quote the time complexity to quantitatively describe the advantage of using $\tilde{U}^\dagger \tilde{U}$. 
For a general full rank matrix of size $4^N \times 4^N$, the time complexity involved in computing 
the inverse operation is $O\left(\left(4^N\right)^3\right)$. 
For comparison, $\tilde{U}^\dagger \tilde{U}$ has non-zero elements only at the indices of
$\left[(k-1)\times2^N+k+l,\, 1+l\right]$ and $\left[1+l,\, (k-1)\times2^N+k+l\right]$, 
where $k = 1,\, 2,\, \ldots,\, 2^N$
and $l$ can be any non-negative integers for the indices to be valid.
The number of non-zero elements in each row of $\tilde{U}^\dagger \tilde{U}$ is less than $2^N$, and thus 
the number of column elementary operations needed for each row is less than $2^N$ during matrix inversion; 
the total number of operations for the $4^N$ rows is less than $2^N \times 2^N \times 4^N$. 
We conclude that the time complexity of solving the inverse matrix of $\tilde{U}^\dagger \tilde{U}$ is $O\left(\left(4^N\right)^2\right)$.

\begin{figure*}[hb]
  \centering
  \includegraphics[width=7in,clip=True]{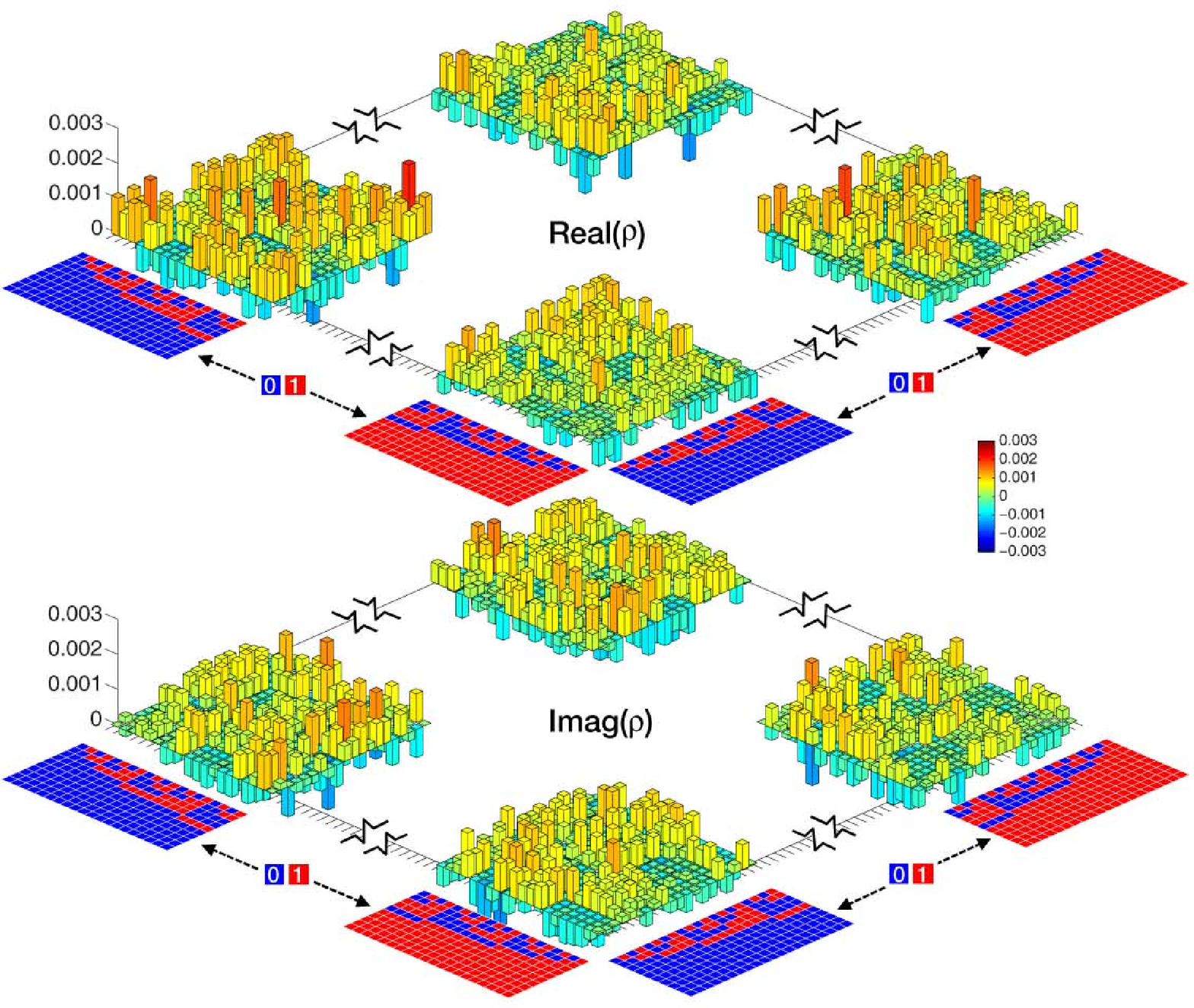}\\
  \caption{\textbf{10-qubit tomography.} 
	Partial matrix elements for the 10-qubit GHZ $\rho$ in the $|0\rangle$ and $|1\rangle$ basis.}\label{fig.rho}
\end{figure*}

\subsection{3.3. 10-qubit $\rho$ in the $|0\rangle$ and $|1\rangle$ basis}
The GHZ density matrix shown in Fig.~3 of the main text has four major elements in the $|\pm\rangle$ basis, while our
measurement is in the $|0\rangle$ and $|1\rangle$ basis. Here we show the partial matrix elements
for the 10-qubit GHZ density matrix in the $|0\rangle$ and $|1\rangle$ basis. It is seen that all matrix elements of $\rho$
are no higher than 0.003 in amplitude (Fig.~\ref{fig.rho}).\\


\begin{thebibliography}{}
\bibitem[S1]{comment} Later on a separate experiment measuring the heating rate of each qubit indicates that
the equilibrium $|1\rangle$-state populations range from 0.2\% to 2\% among 
the 10 qubits on the same chip. The investigation is in progress and the detailed results will be presented elsewhere.
\end{thebibliography}
\end{document}